Analysis of exchange interactions in dimers of $Mn_3$ single-molecule magnets, and their sensitivity to external pressure

Jie-Xiang Yu[1,2,3], George Christou[2,4] and Hai-Ping Cheng[1,2,4*]

[1]Department of Physics, [2]The M$^2$QM Center, [3]The Quantum Theory Project and

[4]Department of Chemistry

Abstract

In light of the potential use of single-molecule magnets (SMMs) in emerging quantum information science initiatives, we report first-principles calculations of the magnetic exchange interactions in $[Mn_3]_2$ dimers of $Mn_3$ SMMs, connected by covalently-attached organic linkers, that have been synthesized and studied experimentally by magnetochemistry and EPR spectroscopy. Energy evaluations calibrated to experimental results give the sign and order of magnitude of the exchange coupling constant ($J_{12}$) between the two $Mn_3$ units that match with fits of magnetic susceptibility data and EPR spectra. Downfolding into the Mn *d*-orbital basis, Wannier function analysis has shown that magnetic interactions can be channeled by ligand groups that are bonded by van der Waals interaction and/or by the linkers via covalent bonding of specific systems, and effective tight-binding Hamiltonians are obtained. We call this long-range coupling that involves a group of atoms a *collective exchange*. Orbital projected spin density of states and alternative Wannier transformations support this observation. To assess the sensitivity of $J_{12}$ to external pressure, stress-strain curves have been investigated for both hydrostatic and uniaxial pressure, which have revealed a switch of $J_{12}$ from ferromagnetic to antiferromagnetic with increasing pressure.

Magnetic molecules have stimulated intense interest in chemistry and physics because of their potential applications in quantum information science [1-4] and other future technologies [5-8]. One important class of magnetic molecules are single-molecule magnets (SMMs), which have a sufficiently large anisotropy barrier vs k$T$ to magnetization relaxation that they behave as nanoscale magnets. One of the most investigated properties in SMMs is quantum tunneling of the magnetization vector (QTM) through the anisotropy barrier[9-13]. In dimers of $Mn_4$ SMMs with $S = {}^9/_2$ held together by hydrogen-bonding, the resulting weak inter-$Mn_4$ exchange coupling led to discovery of new phenomena in molecular magnetism, namely (i) exchange-biased QTM whereby the neighboring $Mn_4$ acted as a bias field shifting the position of the QTM steps in the hysteresis loop[11], and (ii) the generation of quantum superposition and entanglement states of the two $Mn_4$ units, which were identified by analysis of the high-frequency EPR (HF-EPR) spectra[12]. More recently in 2007, triangular $[Mn_3O(O_2CR)_6(py)_3](ClO_4)$ SMMs (R = Me, Et, Ph) were synthesized containing three ferromagnetically-coupled $Mn^{III}$ ions and an $S = 6$ ground state [14]. Covalent linkage of these $Mn_3$ SMMs with organic linkers provides a more controlled method than hydrogen-bonding to synthesize a variety of $[Mn_3]_2$ dimers[14] and higher $[Mn_3]_4$ oligomers [10]. The covalent linkage introduces weak exchange coupling between two $Mn_3$ units in dimers or tetramers, and its sign can be varied depending on the linker identity. These dimers and tetramers were obtained as molecular crystals permitting full structural characterization by X-ray crystallographic studies.

The primary subject of this paper is the $[Mn_3]_2$ dimer with $dpd^{2-}$ linkers ($dpd^{2-}$ = the dianion of 1,3-di(pyridin-2-yl)propane-1,3-dione dioxime), which was found by magnetic susceptibility measurements and HF-EPR spectroscopy to have a ferromagnetic (**FM**) interaction ($J_{12}$) between the two $Mn_3$ units [14]. In addition, HF-EPR was employed to investigate the resulting quantum superposition states in both the crystal and solution phases [14]. The spin Hamiltonian employed was $\hat{H} = \hat{H}_1 + \hat{H}_2 - 2J_{12}\hat{S}_1 \cdot \hat{S}_2$ with $\hat{H}_{1,2} = D\hat{S}_{1,2z}^2 + g\mu_B\hat{S} \cdot \vec{H}$, where $\hat{S}_{1,2}$ and $\hat{H}_{1,2}$ are the spin and the spin Hamiltonian of each $Mn_3$ unit 1 and 2, $\hat{S}_z$ is the spin projection along the easy axis, the anisotropy $D$ = -0.22 cm$^{-1}$, and $g_z$ = 2.00; $\mu_B$ is the Bohr magneton and $\mu_0$ is the vacuum permeability. Both the fits of magnetic and EPR data give $J_{12}$ = 0.025 cm$^{-1}$.

In the absence of theoretical calculations, it was proposed [15] that the Mn $d_\pi$ magnetic orbitals delocalize into the $dpd^{2-}$ $\pi$ systems and the $J_{12}$ exchange coupling is propagated by spin polarization of the bonding electrons in the central $sp^3$ C atom of the linker. Does this explanation give the exact microscopic physics of magnetic exchange in the $[Mn_3]$ dimer? This is the first question we decided to address. The second question is if and how one can tune the $J_{12}$ interaction of a given dimer, and thus the resulting quantum properties. This requires an assessment of the sensitivity of $J_{12}$ to external influences, and an experimentally feasible method is to apply external pressure and see how coupling strength changes. Pressure-dependent investigations have lately become extensively used to uncover new physics behaviors in various materials and states of matter. Particular to magnetic orderings, pressure-dependent structural changes lead to changes in charge and spin ordering, magnetic properties, superconducting transitions [16-17] and transport behavior in perovskites [18-21]. Specific to molecular magnetic materials, pressure- or strain-dependent studies include enhancing the magnetic ordering temperature in rhenium(IV) molecules [22], changing steps in quantum tunneling magnetization [23], and modulating tunneling splitting [24]. It is worthwhile to mention that sometimes 'chemical pressure' (ligand-induced perturbation of magnetic cores) is used to modulate magnetic ordering in molecular magnetic systems via altering the volume of solid unit cells [25-26].

In this letter, we report results from first-principles calculations within the framework of density functional theory [27-28]. We will show below that our investigations provide electronic and magnetic structures that are necessary for understanding long-range *collective magnetic exchange* interaction, test the validity of isolated ions model vs. full-crystal models, and provide strain-stress-magnetism relations. Answers to the two questions posted above will fill gaps in current knowledge of SMM clusters and in theoretical modelling, and guide future experiments and the design of new electronic magnetic materials that are pivotal for next generation electronics and technology for quantum information sciences.

Our DFT-based calculations are performed with projector augmented wave pseudopotentials[29] implemented in the Vienna ab initio simulation [30] (VASP) package. The generalized gradient approximation (GGA) in Perdew, Burke, and Ernzerhof (PBE) formation [28] is used as the exchange-correlation energy and the Hubbard $U$ method [31-32] ($U$ =2.8 eV, $J$ = 0.9 eV) is applied on Mn(3$d$) orbitals to include strong-correlation effects. The magnitude of $U$ is determined by the experimental result for the energy difference between inter-[$Mn_3$] FM and AFM spin configurations per dimer $E_{AF} - E_{FM}$ of about 0.2 meV [see Supplementary Material for details]. An energy cutoff of 600 eV is used for the plane-wave expansion throughout the calculations. The DFT-D3 method with inclusion of van der Waals correction [33] is employed. Total energy, structure optimizing, and electronic structure calculations were performed on three single-molecule [Mn3]-dimers and their ions (we call these gas phase molecules or ions) and one FM dimer in its bulk form. For ions with charge in gas phases, the jellium model is used to keep the whole system charge neutral, and a periodic cubic lattice with lattice constant 24.00~30.00 Å is used to keep molecules sufficiently separated in adjacent unit cells. The $\Gamma$ point only is used for both gas and bulk phases. A $3 \times 3 \times 3$ $\Gamma$-centered $\boldsymbol{k}$-mesh for bulk phases is used for the density-of-state (DOS) calculations. During most structure optimizing calculations, all forces on each atom are less than 1 meV/Å, or 0.2 meV/Å when external strain is applied. With these criteria, the error of the total energy difference between ferromagnetic (FM) and antiferromagnetic (AFM) spin configurations $\delta(E_{AF} - E_{FM})$ is estimated to be about 0.05 meV, which might be reduced further with an energy cutoff larger than 600 eV. After we obtain the eigenstates and eigenvalues, a unitary transformation of Bloch waves is performed to construct the tight-binding Hamiltonian in a Wannier function (WF) basis by using the maximally localized Wannier functions (MLWF) method [34] implemented in the Wannier90 package [35]. The WF-based Hamiltonian has the same exact eigenvalues as those obtained by DFT calculations within an inner energy window, and the outer energy window used for downfolding WF basis is that for all the DFT Bloch states.

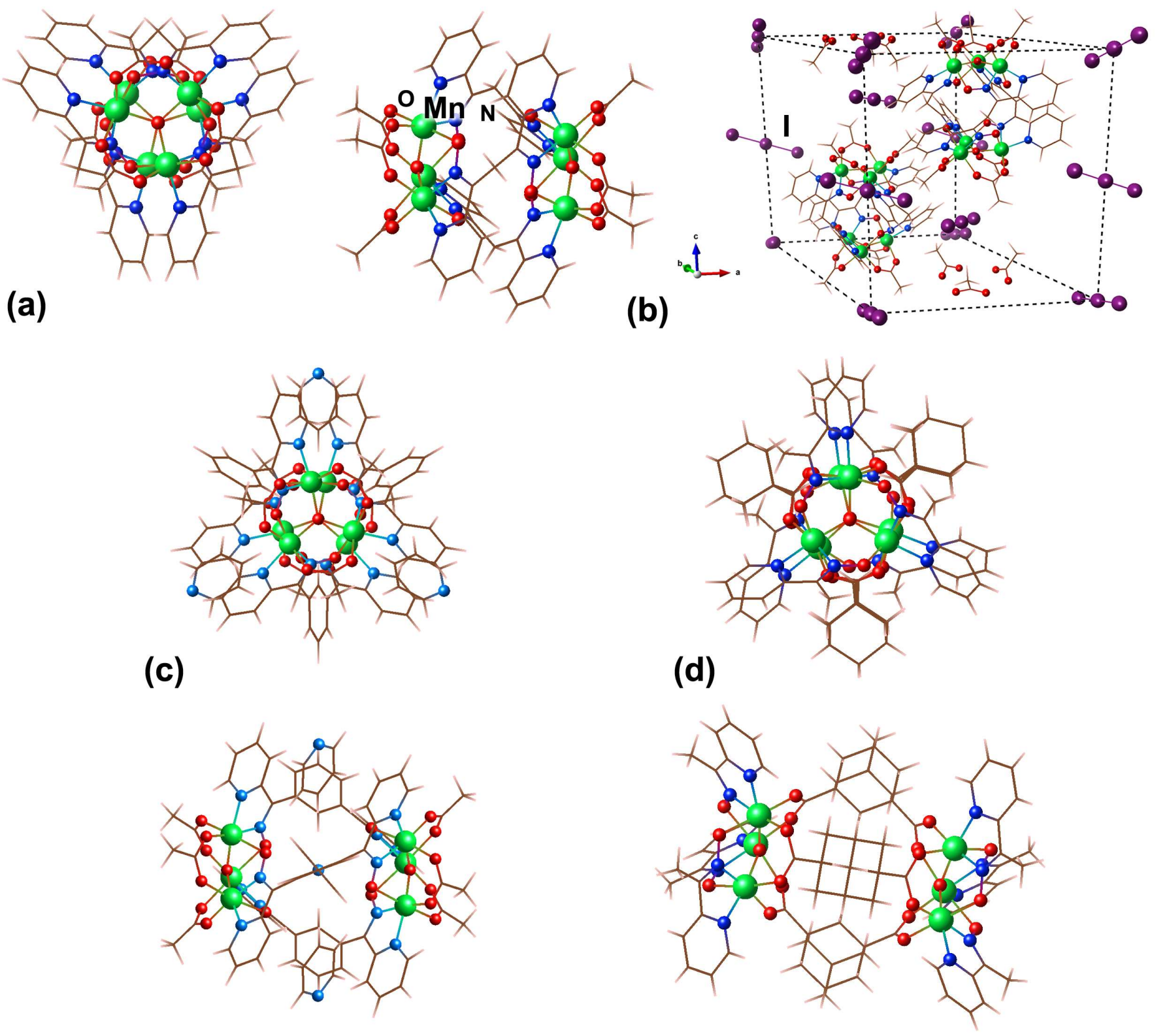

Fig. 1 structure of ($[Mn_3]_2$-$(dpd)_3)^{2+}$, in (a) gas phase and (b) crystal phase. Structure of a (c) ($[Mn_3]_2$-$(ppmd)_3)^{2+}$ and (d) ($[Mn_3]_2$-$(ada)_3)^{2+}$ molecule.

The structures of three SMM $[Mn_3]_2$ dimers are shown in Fig. 1, where two $Mn_3O(O_2CMe)_3$ units are linked covalently by three different linkers. Each $[Mn_3]$ unit has $C_3$ symmetry and a $S = 6$ spin ground state from FM coupling of its three $Mn^{3+}$ ions (each $S$ =2). In the resulting dimer, the two $Mn_3$ triangular planes made are parallel. The dimer with the shortest linker is ($[Mn_3]_2$-$(dpd)_3)^{2+}$ For this system we investigated its crystalline phase as well as individual dimers in isolation. According to experiment [15], the molecular crystal phase of the ($[Mn_3]_2$-$(dpd)_3)^{2+}$ dimer has space group $P\overline{3}\,1c$ with trigonal symmetry in the hexagonal lattice. Each unit cell contains two dimers, which are parallel. The counterions of the ($[Mn_3]_2$-$(dpd)_3)^{2+}$ dimer are two $I_3^-$ anion, so that each unit cell contains four $I_3^-$. Structure optimization was performed with experimental lattice constants $a = 16.569$ Å, $c = 18.285$ Å; Figure 1b shows the optimized atomic structure of ($[Mn_3]_2$- $(dpd)_3)_2(I_3)_4$, even though experimental studies could not determine the location and orientation of the $I_3^-$ counter ions.

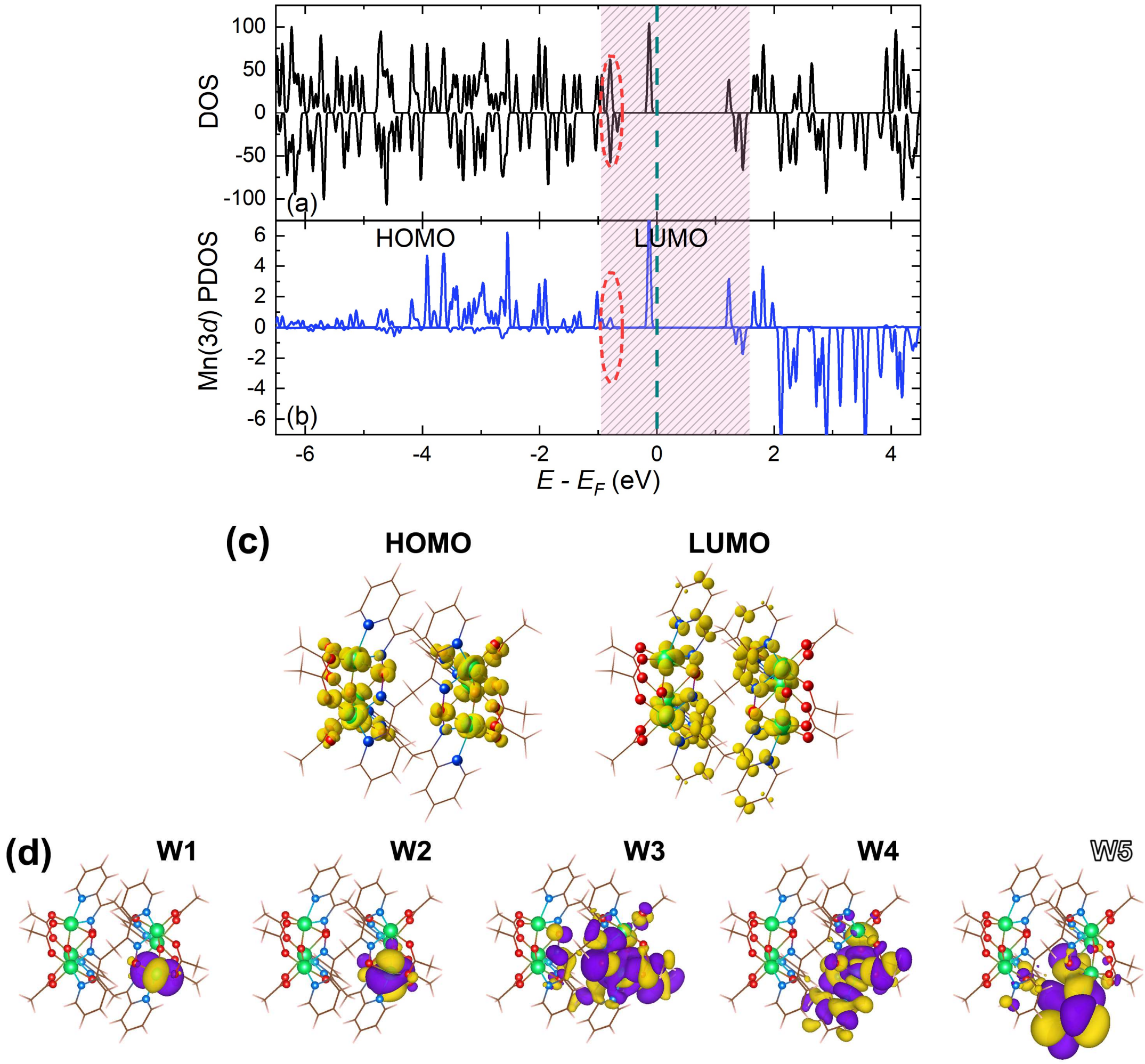

Fig. 2 Electronic structure for the gas phase of $([Mn_3]_2\text{-}(dpd)_3)^{2+}$ dimers. (a) The total density-of-state (DOS) and (b) orbit-resolved projected DOS (PDOS) of Mn(3*d*) orbitals. Positive and negative values correspond to spin-majority and spin-minority channels respectively. The Fermi level is set to zero. The marked state refers to the sub-HOMO state with few Mn(3*d*) components and the shadow region refers to the inner energy window for MLWF. (c) Partial charge density of HOMO (left) and LUMO (right). (d) Aligned by on-site energies, five Wannier functions (WFs) in the spin-majority channel centered on the same Mn atom. Solid and hollow labels represent occupied and unoccupied WFs respectively.

The calculated ground state of the isolated (gas phase) $([Mn_3]_2\text{-}(dpd)_3)^{2+}$ dimer has spin equal to 12, with each $[Mn_3]$ monomer in a $S = 6$ spin state (magnetic moment = $12\mu_B$). Fig. 2a-b shows the total DOS and projected density-of-state (PDOS) of Mn(3*d*) orbitals. The gap between the highest occupied molecular orbital (HOMO) and the lowest unoccupied molecular orbital (LUMO) is 1.2 eV. The isosurfaces of HOMO and LUMO partial charge density shown in Fig. 2c are all in the spin-majority or spin-up channel, and both are dominated by Mn(3$d$) orbitals. However, just below the HOMO, the states circled by red dashed lines (Fig. 2a&b) have few features of Mn(3$d$).

The inter-$Mn_3$ magnetic interaction is weak but long-range. It is best not described as superexchange, which usually involves a single bridging or a few atoms. We therefore denote the magnetic exchange that involves at least one group of atoms as *collective exchange*. To understand the states near the HOMO and LUMO and how they affect the inter-$Mn_3$ exchange interaction, we performed MLWF calculations to get the WF-based Hamiltonian using the inner energy window shown in Fig. 2a&b, with Mn(3$d$) orbitals as the initial projections of Wannier functions for the downfolding process. Fig. 2d shows the five WFs in the spin-majority channel with basis function concentrated on one Mn atom. The two WFs, W1 and W2, with lowest on-site energies about -3.0 eV relative to the Fermi level in the DOS, are just localized on Mn. Three other WFs, W3 to W5 with higher on-site energies, however, have significant contribution from the ligand $O_2CMe$, indicating that the sub-HOMO states are dominated by ligand orbitals. More discussion about ligands is included in the supplementary materials. Considering the $S = 2$ spin state of Mn with four unpaired 3$d$ electrons, W5 with the highest on-site energy is the only unoccupied WF in this WF basis. Other WFs centered on other Mn atoms have similar behaviors, so that W1 to W5 can be regarded as five groups to classify the thirty total WFs in the Hamiltonian. The strongest hopping $t$ in the Hamiltonian between WFs belonging to different sides of the dimer is about 0.070 eV. It comes from one W3-like occupied WF at one side of the dimer and one W5-like unoccupied WF at the other side. Note that the overlap between W3 and W5 is dominated by two acetate ligands (one from each $Mn_3$ unit) instead of linkers and the energy level for the dpd linker is far below HOMO (See Fig.S1a in Supplementary materials). Because the distance between two acetate ligands of different sides is 2.53Å which is in the region of van der Waals interaction, the exchange interaction in this system has major contribution from van der Waals bonded acetate ligands instead of covalent linkers. Magnetic coupling between two van der Waals layers is a subject of great current interest [36]. The electronic structure of the bulk phase $([Mn_3]_2\text{-}(dpd)_3)_2(I_3)_4$ has also been investigated. The FM spin configuration shows that the total spin in one unit cell is $S = 24$ with magnetic moment 48 $\mu_B$, so that each $[Mn_3]$ monomer still has a $S = 6$ spin state with magnetic moment $12.0\mu_B$, indicating the same valence state of the $([Mn_3]_2\text{-}(dpd)_3)^{2+}$ cation as that in the gas phase.

For the other two $[Mn_3]$ dimers, experiments found that $([Mn_3]_2\text{-}(ada)_3)^{2+}$ has a AFM coupling between two $[Mn_3]$ monomers, while the inter-monomer coupling in $([Mn_3]_2\text{-}(ppmd)_3)^{2+}$ is FM. Our total energy calculations for both of them show that $E_{AF} - E_{FM}$ for $([Mn_3]_2\text{-}(ada)_3)^{2+}$ is -0.02 meV per dimer and that for $([Mn_3]_2\text{-}(ppmd)_3)^{2+}$ is 0.03 meV per dimer. These results for $|E_{AF} - E_{FM}|$ are one order smaller than that for $([Mn_3]_2\text{-}(dpd)_3)^{2+}$ and are within our numerical error. However, we can still obtain insight into to electronic structure and exchange pathways. The

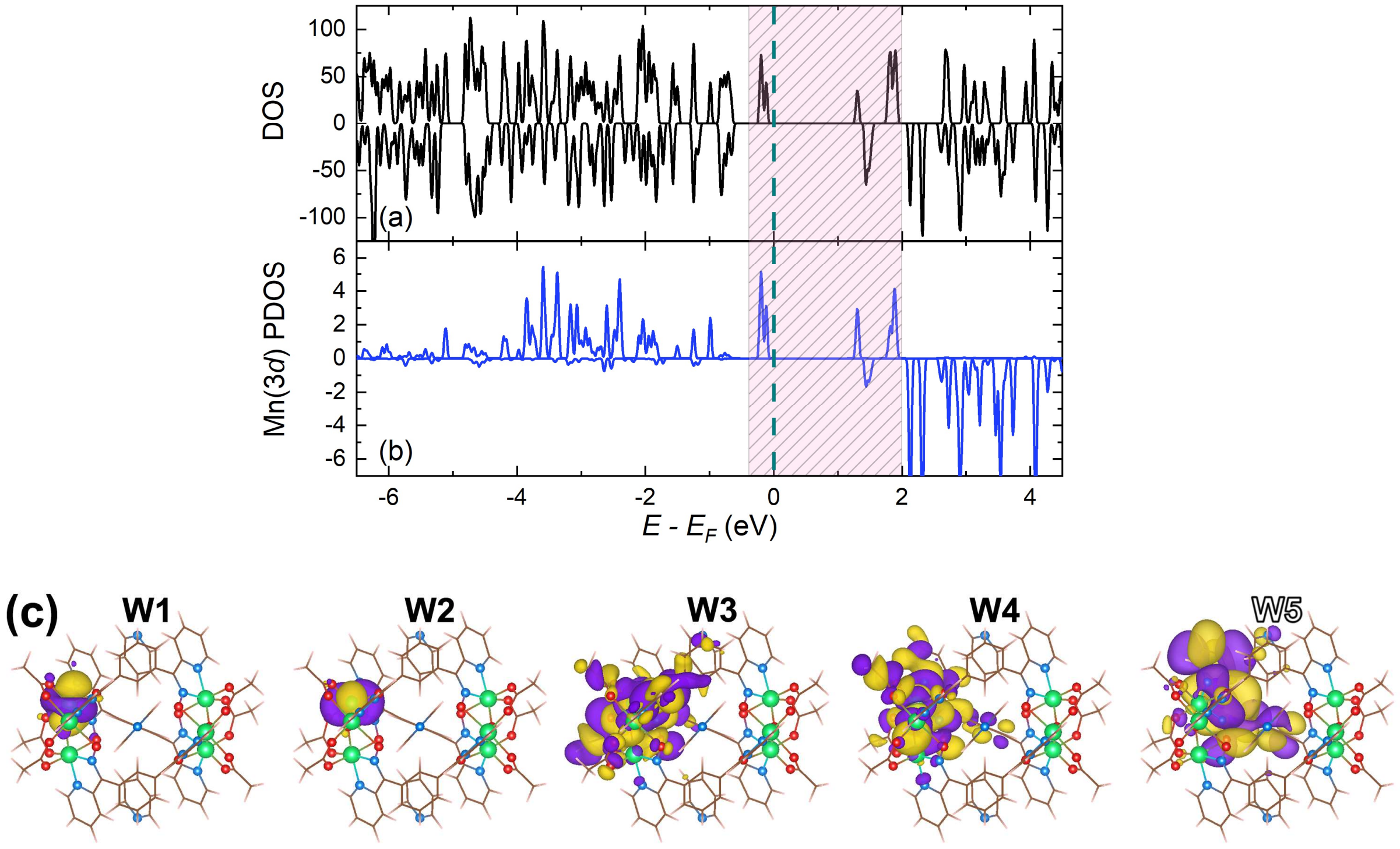

Fig. 3 Electronic structure for the gas phase of $([Mn_3]_2\text{-}(ppmd)_3)^{2+}$ dimers. (a) The total DOS and (b) PDOS of Mn(3*d*) orbitals. The shadow region refers to the inner energy window for MLWF. (c) Ordered by on-site energies, five WFs in the spin-majority channel centered on the same Mn atom. Solid/hollow labels represent occupied and unoccupied WFs respectively.

tiny inter-Mn3 exchange can be explained by the electronic structure. Fig. 3a-b shows the total DOS and PDOS of Mn(3*d*) orbitals for $([Mn_3]_2\text{-}(ppmd)_3)^{2+}$ dimers. The DOS and PDOS are very similar to that for $([Mn_3]_2\text{-}(dpd)_3)^{2+}$, where the HOMO and LUMO are dominated by Mn(3*d*) orbitals and the sub-HOMOs do not have features of Mn(3*d*). The atomic-centered WFs for the tight-binding Hamiltonian for $([Mn_3]_2\text{-}(ppmd)_3)^{2+}$ [Fig. 3c] are also similar to those in $([Mn_3]_2\text{-}(dpd)_3)^{2+}$, where two localized WFs have lower on-site energies, and three other WFs with higher on-site energies have some features of the acetate ligand $O_2CMe$ but mostly the linker ppmd, indicating that the linkers are responsible for the exchange couplings In the WF basis for $([Mn_3]_2\text{-}(ppmd)_3)^{2+}$, W5 with the highest on-site energy is unoccupied. Among all the inter-$Mn_3$ hopping in the Hamiltonian, the strongest hopping $t$ between occupied and unoccupied WFs is -0.016 eV, which again comes from the hopping between the occupied W3-like WF and the unoccupied W5-like WF. Thus, $([Mn_3]_2\text{-}(ppmd)_3)^{2+}$ has much weaker inter-$Mn_3$ hopping. Linker ppmd with a larger size than dpd makes both the hopping crossing the linker and the exchange through the van der

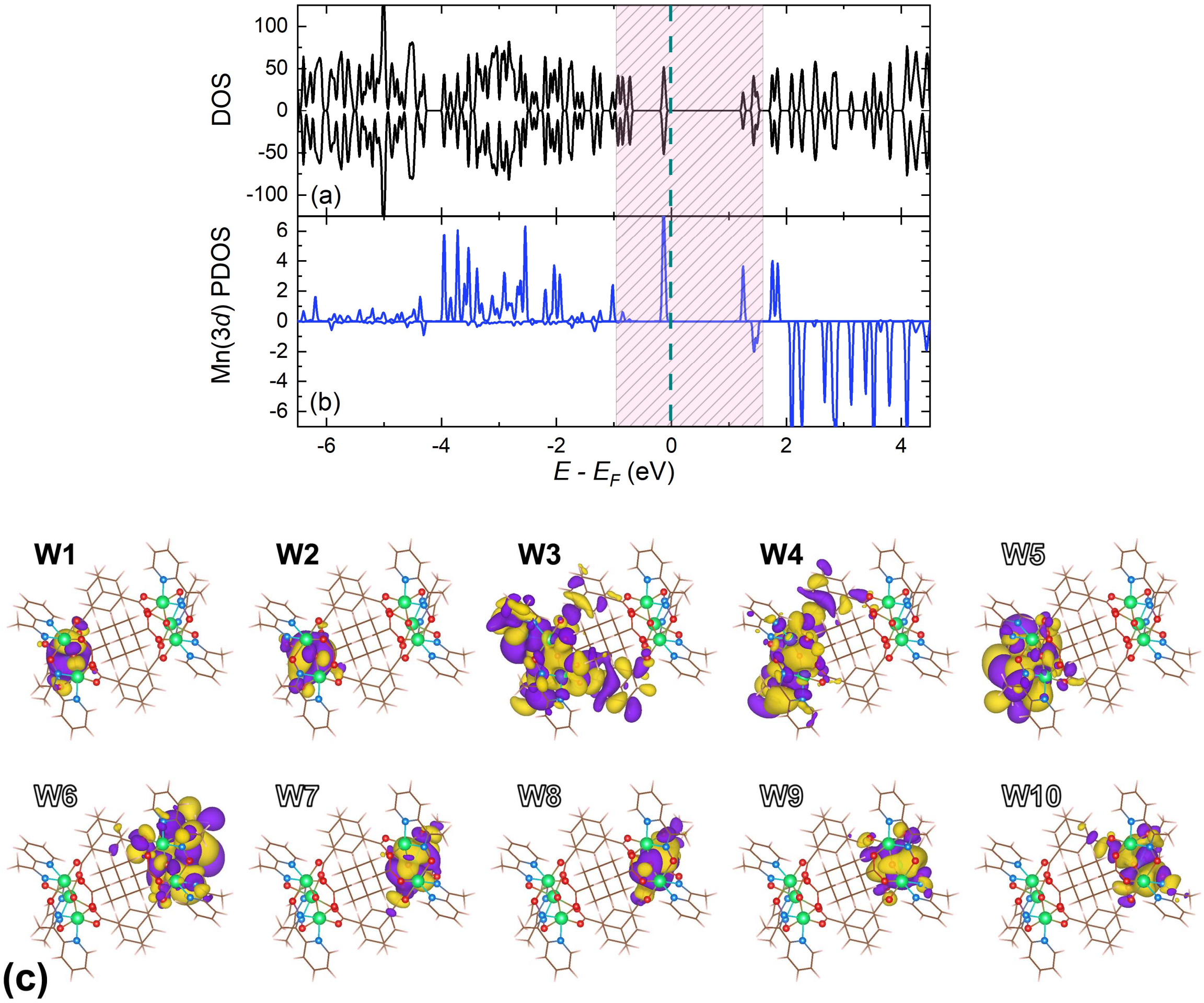

Fig. 4 Electronic structure for the gas phase of $([Mn_3]_2\text{-}(ada)_3)^{2+}$ dimers with AFM spin ordering. (a) The total DOS and (b) PDOS of Mn(3*d*) orbitals. The shadow region refers to the inner energy window for MLWF. (c) In the spin-majority channel, ten WFs, of which five are centered on an Mn atom at one side of the dimer and the other five centered on the corresponding Mn atom at the other side. These WFs are ordered by on-site energies, and solid/hollow labels represent occupied/unoccupied WFs respectively.

Waals bonded ligands weak. Numerically, the strength of the exchange interaction, which is usually proportional to $t^2$ for $([Mn_3]_2\text{-}(ppmd)_3)^{2+}$ is about one order smaller than that for $([Mn_3]_2\text{-}(dpd)_3)^{2+}$, consistent with the total energy results.

The DOS results for AFM $([Mn_3]_2\text{-}(ada)_3)^{2+}$ are shown in Fig. 4a&b. Due to AFM spin-ordering, the total DOS are identical in both spin channels. The PDOS on one Mn still gives the $S = 2$ high-spin state, the same with that for the other two dimers. The atomic-centered WFs for the tight-binding Hamiltonian are also obtained. Note that for this dimer, the two $Mn_3$ units are bonded only through covalent linkers. In the same spin channel, Fig. 4c gives ten WFs, with five of them, W1 to W5, centered on a Mn atom at one side of the dimer and the other five, W6 to W10, centered on the corresponding Mn atom at the other side. That means W1 to W5 are the spin-up atomic-like orbitals of one Mn and W6 to W10 are the spin-down atomic-like orbitals of another. Because of $S$ = 2 high-spin state of Mn with four unpaired $3d$ electrons, the four WFs, W1 to W4, with the lowest energies are occupied in this tight-binding basis and others are unoccupied. Among the occupied WFs, W1 and W2 are localized on Mn while W3 and W4 has a comparatively significant contribution from the ada linkers. Among the unoccupied WFs at the other side of the dimer, all WFs except for W6 are localized on Mn. The strongest inter-monomer hopping $t$ in the Hamiltonian between occupied and unoccupied WFs are -0.015 eV between W3 and W10. The value of $t^2$ is only one twentieth of that in $([Mn_3]_2\text{-}(dpd)_3)^{2+}$. Considering that the only difference between $([Mn_3]_2\text{-}(dpd)_3)^{2+}$ and $([Mn_3]_2\text{-}(ada)_3)^{2+}$ is the different linkers, which affect the strength of inter-$Mn_3$ hopping, we can estimate that the strength of the exchange interaction of $([Mn_3]_2\text{-}(ada)_3)^{2+}$ is one twentieth in magnitude of that in $([Mn_3]_2\text{-}(dpd)_3)^{2+}$, numerically consistent with the total energy results.

Although HF-EPR experiments [15] have shown that covalently-linked $[Mn_3]_2$ dimers remain intact in the solution phase and retain their J12 exchange coupling and resulting quantum superposition states, their low Young's modulus in the solid phase suggests a non-trival strain-sensitive response. To study the strain response of the J12 magnetic interaction, we applied external strain on $([Mn_3]_2\text{-}(dpd)_3)_2(I_3)_4$ by changing the lattice constant. Fig. 5a&b shows the isotropic strain-dependent results. The value of $E_{AF} - E_{FM}$ as a function of strain shows that the magnetic interaction can transition from FM to AFM when a compressive strain of about 0.4% in length, or 1.2% in volume, is applied. The hydrostatic pressure corresponding to this transition is about 0.06 GPa according to Fig. 5b. This strain-induced FM-AFM transition can also happen with a uniaxial strain by changing the out-of-plane lattice constant $c$ and keeping the volume of the unit cell invariant. According to the result shown in Fig. 5d&e, the sign of $E_{AF} - E_{FM}$ shifts from positive to negative when $c$ is changed by about 1.4%; the corresponding compressive pressure in that direction is also about 0.06 GPa.

Since the $[Mn_3]$-dimer is linked covalently by the linker $dpd^{2-}$, the mechanism of the magnetic interaction is dominated by *collective exchange*, which is sensitive to both bond length and bond angle. Fig. 5c shows the distance between two $[Mn_3]$ planes and the C-C-C bond angle of the linker dpd as a function of strain. As both isotropic and uniaxial compressive strain increases, the magnitude of the bond angle and the distance decreases with an almost linear relation. At the transition point, the distance between two $[Mn_3]$ planes is 6.198 Å under isotropic strain, 0.14% closer than the zero pressure value. The number under uniaxial strain is 6.192 Å, 0.24% closer than the zero-pressure case. On the other hand, the C-C-C bond angle of the linker dpd is 112.94°, 0.13% smaller than for zero pressure. The number under uniaxial strain is 112.99°, 0.09% smaller than zero pressure. The percentage of the change of both distances and bond angles are much smaller than the percentage of the applied strain, indicating that the major response to strain is the changing of inter-molecular distances instead of the modification of the bond lengths and angles within one [$Mn_3$]-dimer molecule. However, this small change is enough to drive the FM-AFM transition, reflecting the sensitivity of the response to the magnetic exchange interaction and confirming the collective-exchange feature.

In summary, we have studied the electronic and magnetic properties of a series of SMM [$Mn_3$]$_2$

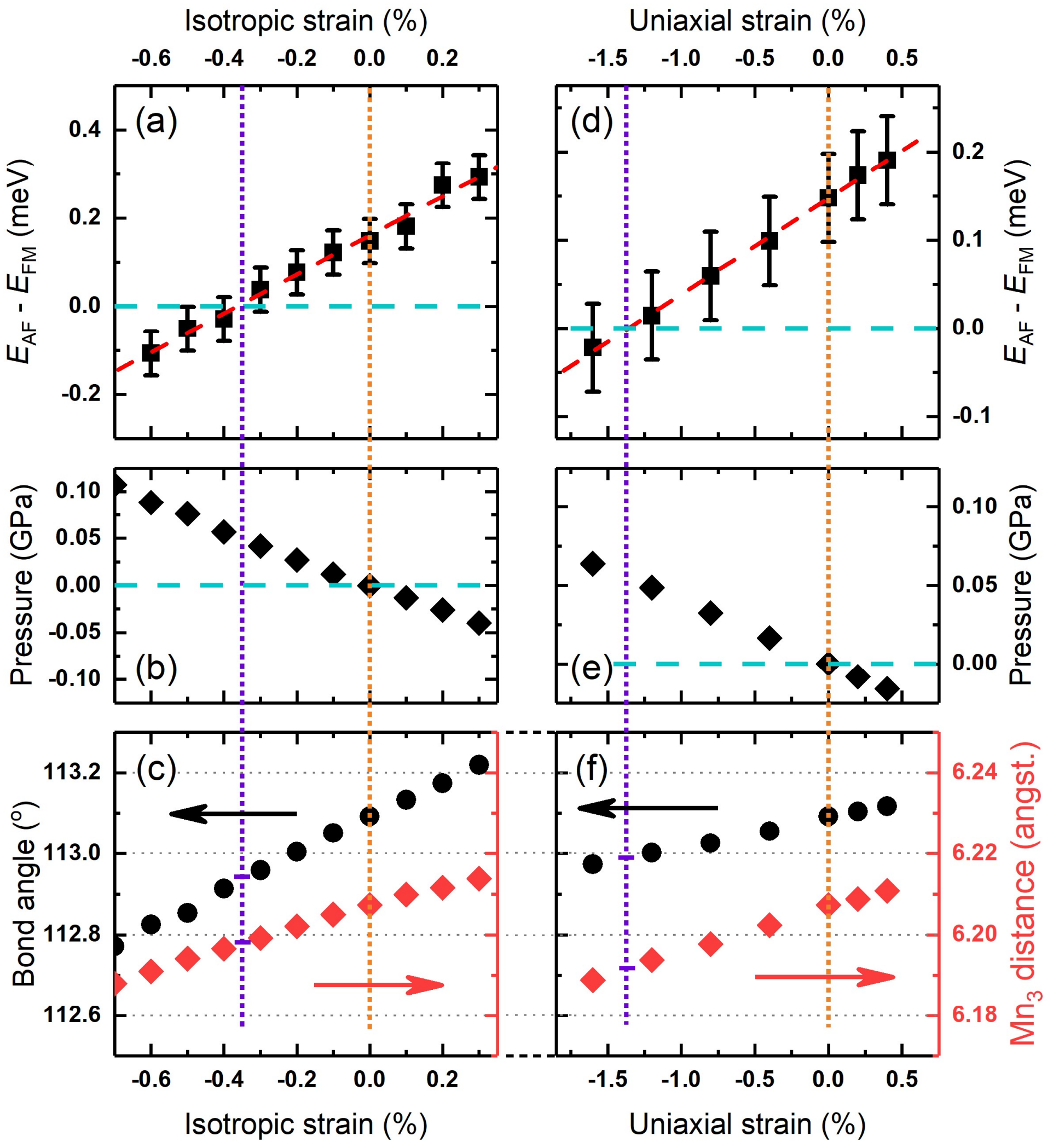

Fig. 5 Strain dependences for ([$Mn_3$]$_2$-(dpd)$_3$)$_2$($I_3$)$_4$ bulk. (a) Energy difference $E_{AF} - E_{FM}$ per dimer, (b) pressure, (c) C-C-C bond angle (black dots) of the linker (dpd) and distance (red dots) between two [$Mn_3$] as a function of isotropic strain. (d), (e), (f) Similar results as a function of uniaxial strain. (c) and (f) share the same vertical scales. Violet and orange vertical dotted lines represent FM/AF transitions and the unstrained results respectively.

dimers based on first-principles calculations. While energetic determination of the magnetic order of dimers relies on experiments due to the weak coupling nature, the coupling pathway can be uncovered by Wannier orbital analysis. Total DOS and PDOS results show similar electronic structures of Mn(3*d*) orbitals among the three dimers. By downfolding into a Mn-centered WF basis, the tight-binding Hamiltonian indicates that only $([Mn_3]_2\text{-}(dpd)_3)^{2+}$ has significant inter-$Mn_3$ hopping via $O_2CMe$ ligands sufficient to bring about FM coupling. The other two $[Mn_3]$-dimers, $([Mn_3]_2\text{-}(ppmd)_3)^{2+}$ and $([Mn_3]_2\text{-}(ada)_3)^{2+}$, have small hopping coefficients, via covalently bonded linkers, leading to a much weaker exchange interaction. These results suggest that, like linkers, van der Waals bonded ligands can also affect the sign and strength of magnetic exchange. The calculated results also show that even though the isolated dimer model is different in structure from the bulk form, the trend in the relative energies and densities of states are the same. It is a reasonable model for Wannier function analysis. Finally, the strain dependence of magnetic interactions in the bulk phase of $([Mn_3]_2\text{-}(dpd)_3)_2(I_3)_4$ are investigated under both hydrostatic and uniaxial pressure. A strain-induced FM-AFM transition is identified when an external pressure of about 0.06 GPa is applied, indicating the sensitivity of the response to magnetic exchange interaction and providing potential application to molecular-based spintronics. Experimental work on pressure dependence is underway.

*hping@ufl.edu

**Acknowledgment:** This work was supported as part of the Center for Molecular Magnetic Quantum Materials, an Energy Frontier Research Center funded by the U.S. Department of Energy, Office of Science, Basic Energy Sciences under Award No. DE-SC0019330. Computations were performed at NERSC and UFRC.

# Supplementary Materials for "Analysis of exchange interactions in dimers of Mn3 single-molecule magnets, and their sensitivity to external pressure"

Jie-Xiang Yu[1,2,3], George Christou[2,4] and Hai-Ping Cheng[1,2,4*]

[1]Department of Physics, [2]The M[2]QM Center, [3]The Quantum Theory Project and [4]Department of Chemistry, University of Florida, Gainesville, FL, USA

## 1. Hubbard *U*-dependent calculations

The inter-$[Mn_3]$ magnetic interaction in the $([Mn_3]_2\text{-}(dpd)_3)^{2+}$ dimer is a very weak inter-unit ferromagnetic coupling, with $J \approx 0.025$ cm$^{-1}$ according to EPR measurements [Ref. 15 in the main text]. Considering spin states $S$ = 6 for each $[Mn_3]$ monomer, we know that the energy difference $E_{AF} - E_{FM}$ between inter-$[Mn_3]$ FM and AFM spin configurations per dimer is about 0.2 meV. Since the exchange coupling is usually highly influenced by Hubbard *U*, it is necessary to choose a reasonable value of Hubbard *U*. Therefore, we calculate $E_{AF} - E_{FM}$ as a function of *U* for a $[Mn_3]_2$ crystal. Without *U*, $E_{AF} - E_{FM} = -2.48$ meV; $E_{AF} - E_{FM} = -0.37$ meV at *U* = 2.5 eV; $E_{AF} - E_{FM} = 0.41$ meV at *U* = 3.0 eV; $E_{AF} - E_{FM} = 1.65$ meV at *U* = 4.0 eV. In this case, $E_{AF} - E_{FM} = 0.13$ meV at *U* = 2.8 eV best matches the experimental result and we use this *U* value in the rest of our calculations.

## 2. Downfolding and Mn-centered WF-based Hamiltonian

As mentioned in the main text, we use the MLWF method to get the WF-based Hamiltonian with Mn(3*d*) orbitals as the initial projections of Wannier functions for the downfolding process. As a result, thirty WFs are obtained, where five WFs are centered on the same Mn. We focus on the inter-$Mn_3$ hopping of the Hamiltonian between two sets of five WFs, one centered on the Mn on the left side of the dimer and the other centered on the Mn on the right side.

**$([Mn_3]_2\text{-}(dpd)_3)^{2+}$**: the inter-$Mn_3$ matrix element or hopping coefficient $t_{lr}$ between Mn-centered WF in the spin-majority channel, in units of eV:

| $t_{lr}$ | $\|W1_r^o\rangle$ | $\|W2_r^o\rangle$ | $\|W3_r^o\rangle$ | $\|W4_r^o\rangle$ | $\|W5_r^u\rangle$ |
|---|---|---|---|---|---|
| $\langle W1_l^o\|$ | −0.005 | 0.004 | −0.032 | −0.011 | **−0.033** |
| $\langle W2_l^o\|$ | −0.004 | −0.004 | −0.047 | −0.011 | **−0.001** |
| $\langle W3_l^o\|$ | −0.032 | 0.044 | 0.003 | −0.003 | **0.070** |
| $\langle W4_l^o\|$ | 0.012 | −0.004 | −0.002 | −0.001 | **−0.005** |
| $\langle W5_l^u\|$ | **−0.032** | **−0.008** | **0.068** | **0.011** | 0.043 |

Labels W1 to W5 are the same as in the main text. Subscripts *l* and *r* represent the left and right side of the dimer respectively. Superscripts *o* and *u* represent occupied and unoccupied WFs respectively. Bold numbers are the hopping between occupied and unoccupied WFs, and red shows the strongest ones. The corresponding full Hamiltonian matrix appears at the end of the document.

Similar matrix elements for $([Mn_3]_2\text{-}(ppmd)_3)^{2+}$ are given by:

| $t_{lr}$ | $\|W1_r^o\rangle$ | $\|W2_r^o\rangle$ | $\|W3_r^o\rangle$ | $\|W4_r^o\rangle$ | $\|W5_r^u\rangle$ |
|---|---|---|---|---|---|
| $\langle W1_l^o\|$ | 0.000 | 0.000 | 0.000 | −0.002 | **0.002** |
| $\langle W2_l^o\|$ | 0.000 | 0.000 | −0.006 | −0.001 | **0.002** |
| $\langle W3_l^o\|$ | 0.004 | −0.004 | 0.008 | −0.006 | **−0.018** |
| $\langle W4_l^o\|$ | −0.002 | 0.000 | 0.006 | 0.002 | **0.005** |
| $\langle W5_l^u\|$ | **−0.003** | **0.000** | **−0.019** | **−0.004** | −0.016 |

The corresponding full Hamiltonian matrix appears at the end of the document.

For $([Mn_3]_2\text{-}(ada)_3)^{2+}$ in the AFM spin configuration, we let Mn on the left side of the dimer (Fig. 4(c) in the main text) be spin-up, so that in the spin-majority channel, four of five Mn-centered WFs are occupied on the left side while all Mn-center WFs are unoccupied at the right side. Thus, the corresponding inter-$Mn_3$ hopping $t_{lr}$ is given by:

| $t_{lr}$ | $\|W1_r^u\rangle$ | $\|W2_r^u\rangle$ | $\|W3_r^u\rangle$ | $\|W4_r^u\rangle$ | $\|W5_r^u\rangle$ |
|---|---|---|---|---|---|
| $\langle W1_l^o\|$ | **0.001** | **0.001** | **0.000** | **−0.001** | **0.000** |
| $\langle W2_l^o\|$ | **−0.001** | **0.001** | **0.000** | **0.000** | **−0.001** |
| $\langle W3_l^o\|$ | **−0.002** | **0.003** | **0.001** | **−0.002** | **−0.015** |
| $\langle W4_l^o\|$ | **−0.004** | **0.002** | **0.001** | **0.005** | **−0.003** |
| $\langle W5_l^u\|$ | 0.001 | 0.001 | −0.001 | −0.001 | 0.000 |

Again, the corresponding full Hamiltonian matrix appears at the end of the document.

## 3. Projected density-of-state of linkers

The PDOS of the linkers of three $[Mn_3]$-dimers are shown in Fig. S1. For $([Mn_3]_2\text{-}(dpd)_3)^{2+}$, the component of dpd are mainly at -6~-4 eV which is far from the Femi level. On the other hand, the component of ppmd in $([Mn_3]_2\text{-}(ppmd)_3)^{2+}$ and that of ada in $([Mn_3]_2\text{-}(ada)_3)^{2+}$ has significant features at -2~-1 eV which is near the Fermi level. To this end, the linker dpd are not involved into the exchange pathway while the linkers ppmd and ada play key roles on exchange interaction.

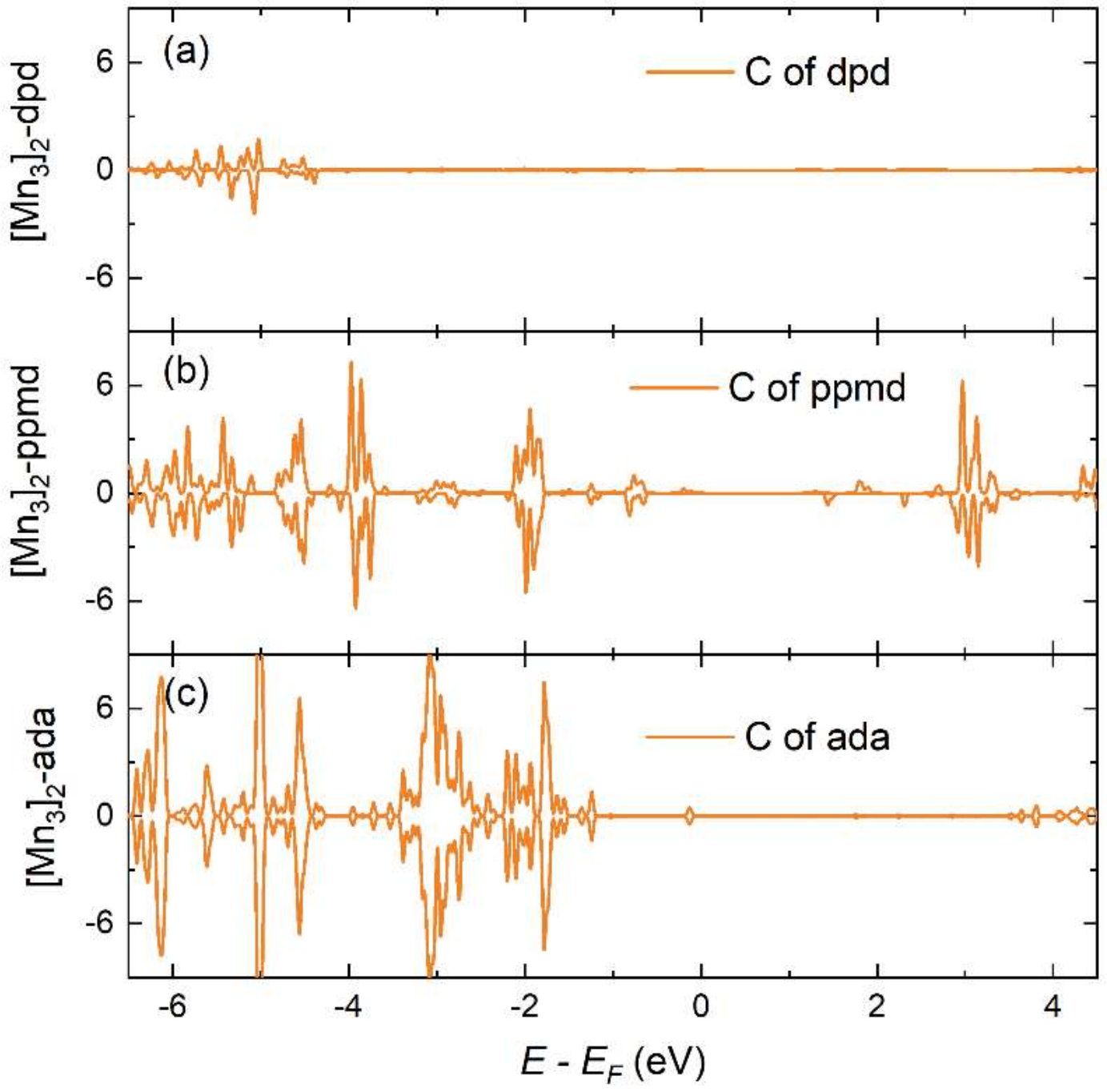

Fig. S1 In their gas phases, the PDOS of 2s and 2p orbitals of (a) the carbon of dpd, (b) six carbons of ppmd and (c) ten carbons of ada.

# 4. Alternative WF-based Hamiltonian near the Fermi level

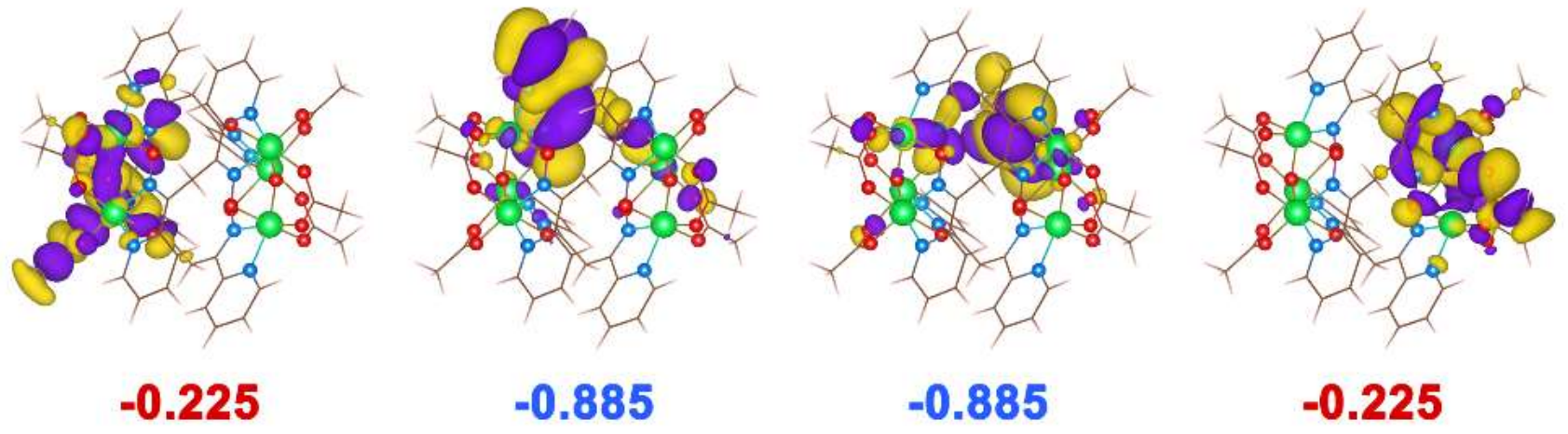

Fig. S2 Four of the WFs obtained from the top fourteen occupied eigenstates in $([Mn_3]_2\text{-}(dpd)_3)^{2+}$ near the Fermi level. Red and blue numbers represent the on-site energies relative to the Fermi level in units of eV for two Mn-centered and two $O_2CMe$ ligand-dominated WFs respectively.

To see clearly what local orbitals are associated with the sub-HOMO states, that is, the hybridization of ligand orbitals and Mn(3*d*) orbitals that dominate HOMO/LUMO levels in dimer systems, circled by green dashed line in Fig. 2 in the main text, we use a gas phase model $([Mn_3]_2\text{-}(dpd)_3)^{2+}$ to perform a unitary transformation on Bloch waves near the Fermi level. This time, we use both Mn(3d) and ligand orbitals as base functions without downfolding. The eigenvalues of the WF-based tight-binding Hamiltonian can reproduce the eigenvalues obtained by DFT.

The top fourteen occupied eigenstates in the spin-majority channel are included in the energy window. As a result, fourteen corresponding MLWFs are obtained. Fig. S2 shows four typical WFs. Two of them, with lower on-site energies, display significant features of $O_2CMe$ ligands, indicating that the sub-HOMO states are dominated by $O_2CMe$ ligands. Compared to the Mn(3*d*)-only WF construction which gives an *effective* Hamiltonian for long-range *collective-exchange,* this Wannier transformation demonstrates the physical processes and channels for exchange coupling between the two [$Mn_3$] monomers.

## 5. Density-of-state results for ([$Mn_3$]$_2$-(dpd)$_3$)$_2$($I_3$)$_4$ bulk

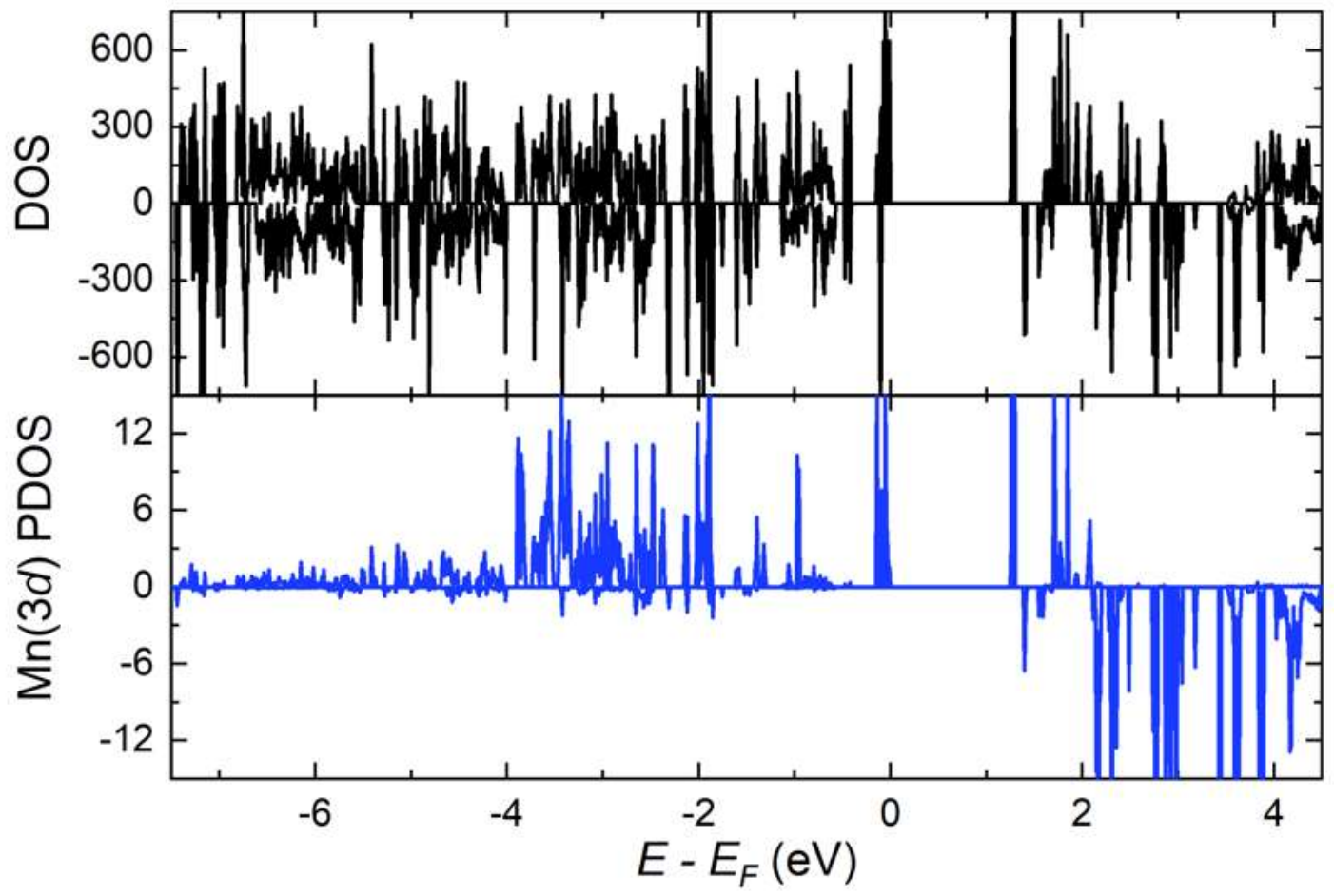

Fig. S3 the total DOS and PDOS of Mn(3*d*) orbitals in the bulk phase of ([$Mn_3$]$_2$-(dpd)$_3$)$_2$($I_3$)$_4$.

DOS and PDOS in the bulk phase of ([$Mn_3$]$_2$-(dpd)$_3$)$_2$($I_3$)$_4$ (shown in Fig. S3) gives very the same electronic structure as that in the gas phase.

**Appendix: Tight-binding Hamiltonian**

The matrix elements of the three Mn-centered tight-binding Hamiltonians are listed below in three sets of columns. In each set, the first and second column are the indices of the WFs and the third column is the matrix element of the Hamiltonian. For $([Mn_3]_2\text{-}(dpd)_3)^{2+}$ and $([Mn_3]_2\text{-}(ppmd)_3)^{2+}$, orbitals 1~5, 6~10, and 11~15 are three sets of WFs centered on three Mn atoms on the left side of the dimer; 16~20, 21~25, 26~30 are three sets of WFs centered on three Mn atoms on the right side. For $([Mn_3]_2\text{-}(ada)_3)^{2+}$, orbitals 1~5, 11~15, and 21~25 are WFs centered on the three spin-up Mn atoms at the left side, and 6~10, 16~20, 26~30 are centered on the three spin-down Mn atoms at the right side. WFs with similar on-site energies are classified into the same label. For example, Nos. 1, 10, 11, 16, 21 and 30 in $([Mn_3]_2\text{-}(dpd)_3)^{2+}$ have the lowest on-site energies among their five WFs centered on the corresponding Mn, so they are all labeled by WF1 in the main text.

**Tight-binding Hamiltonian of $([Mn_3]_2\text{-}(dpd)_3)^{2+}$**

| | | | | | | | | |
|---|---|---|---|---|---|---|---|---|
| 1 | 1 | -8.225 | 1 | 11 | -0.014 | 1 | 21 | 0.002 |
| 2 | 1 | 0.190 | 2 | 11 | -0.008 | 2 | 21 | 0.008 |
| 3 | 1 | 0.050 | 3 | 11 | -0.086 | 3 | 21 | 0.034 |
| 4 | 1 | -0.052 | 4 | 11 | -0.173 | 4 | 21 | 0.005 |
| 5 | 1 | -0.077 | 5 | 11 | 0.066 | 5 | 21 | 0.001 |
| 6 | 1 | -0.056 | 6 | 11 | -0.028 | 6 | 21 | 0.006 |
| 7 | 1 | 0.061 | 7 | 11 | -0.037 | 7 | 21 | 0.036 |
| 8 | 1 | -0.010 | 8 | 11 | -0.127 | 8 | 21 | 0.035 |
| 9 | 1 | 0.088 | 9 | 11 | -0.263 | 9 | 21 | -0.020 |
| 10 | 1 | -0.031 | 10 | 11 | -0.030 | 10 | 21 | 0.003 |
| 11 | 1 | -0.014 | 11 | 11 | -8.265 | 11 | 21 | -0.005 |
| 12 | 1 | 0.130 | 12 | 11 | 0.030 | 12 | 21 | 0.032 |
| 13 | 1 | -0.039 | 13 | 11 | -0.109 | 13 | 21 | 0.033 |
| 14 | 1 | -0.035 | 14 | 11 | 0.043 | 14 | 21 | -0.004 |
| 15 | 1 | 0.175 | 15 | 11 | -0.152 | 15 | 21 | -0.012 |
| 16 | 1 | -0.005 | 16 | 11 | -0.001 | 16 | 21 | -0.014 |
| 17 | 1 | -0.032 | 17 | 11 | -0.023 | 17 | 21 | 0.039 |
| 18 | 1 | -0.032 | 18 | 11 | 0.029 | 18 | 21 | -0.130 |
| 19 | 1 | -0.004 | 19 | 11 | 0.004 | 19 | 21 | -0.034 |
| 20 | 1 | 0.012 | 20 | 11 | -0.016 | 20 | 21 | -0.173 |
| 21 | 1 | 0.002 | 21 | 11 | -0.005 | 21 | 21 | -8.228 |
| 22 | 1 | 0.008 | 22 | 11 | -0.033 | 22 | 21 | 0.192 |
| 23 | 1 | -0.035 | 23 | 11 | 0.016 | 23 | 21 | -0.048 |
| 24 | 1 | 0.005 | 24 | 11 | -0.011 | 24 | 21 | -0.051 |
| 25 | 1 | -0.001 | 25 | 11 | -0.004 | 25 | 21 | 0.077 |
| 26 | 1 | 0.006 | 26 | 11 | 0.002 | 26 | 21 | -0.056 |
| 27 | 1 | 0.036 | 27 | 11 | -0.030 | 27 | 21 | 0.060 |
| 28 | 1 | -0.035 | 28 | 11 | -0.008 | 28 | 21 | 0.010 |
| 29 | 1 | -0.020 | 29 | 11 | -0.004 | 29 | 21 | 0.087 |
| 30 | 1 | -0.003 | 30 | 11 | 0.001 | 30 | 21 | 0.031 |
| 1 | 2 | 0.190 | 1 | 12 | 0.130 | 1 | 22 | 0.008 |
| 2 | 2 | -3.517 | 2 | 12 | -0.013 | 2 | 22 | -0.002 |
| 3 | 2 | -0.006 | 3 | 12 | 0.042 | 3 | 22 | -0.001 |
| 4 | 2 | 0.025 | 4 | 12 | 0.026 | 4 | 22 | -0.004 |
| 5 | 2 | -0.674 | 5 | 12 | 0.012 | 5 | 22 | 0.002 |
| 6 | 2 | -0.096 | 6 | 12 | 0.009 | 6 | 22 | 0.032 |
| 7 | 2 | 0.045 | 7 | 12 | 0.037 | 7 | 22 | -0.040 |
| 8 | 2 | 0.014 | 8 | 12 | -0.013 | 8 | 22 | -0.044 |
| 9 | 2 | 0.026 | 9 | 12 | -0.014 | 9 | 22 | 0.066 |
| 10 | 2 | 0.089 | 10 | 12 | -0.005 | 10 | 22 | -0.013 |
| 11 | 2 | -0.008 | 11 | 12 | 0.030 | 11 | 22 | -0.033 |
| 12 | 2 | -0.013 | 12 | 12 | -3.513 | 12 | 22 | -0.043 |
| 13 | 2 | -0.037 | 13 | 12 | 0.001 | 13 | 22 | -0.070 |
| 14 | 2 | -0.004 | 14 | 12 | -0.670 | 14 | 22 | -0.001 |
| 15 | 2 | -0.016 | 15 | 12 | -0.024 | 15 | 22 | 0.005 |
| 16 | 2 | -0.033 | 16 | 12 | -0.029 | 16 | 22 | -0.008 |
| 17 | 2 | 0.070 | 17 | 12 | -0.035 | 17 | 22 | 0.037 |
| 18 | 2 | 0.043 | 18 | 12 | 0.043 | 18 | 22 | 0.013 |
| 19 | 2 | -0.001 | 19 | 12 | -0.021 | 19 | 22 | -0.004 |
| 20 | 2 | -0.005 | 20 | 12 | -0.068 | 20 | 22 | 0.016 |
| 21 | 2 | 0.008 | 21 | 12 | 0.032 | 21 | 22 | 0.192 |
| 22 | 2 | -0.002 | 22 | 12 | -0.043 | 22 | 22 | -3.520 |
| 23 | 2 | 0.001 | 23 | 12 | 0.068 | 23 | 22 | 0.005 |
| 24 | 2 | -0.004 | 24 | 12 | -0.011 | 24 | 22 | 0.025 |
| 25 | 2 | -0.002 | 25 | 12 | -0.008 | 25 | 22 | 0.673 |
| 26 | 2 | 0.032 | 26 | 12 | -0.005 | 26 | 22 | -0.096 |
| 27 | 2 | -0.041 | 27 | 12 | -0.001 | 27 | 22 | 0.045 |
| 28 | 2 | 0.044 | 28 | 12 | -0.002 | 28 | 22 | -0.014 |
| 29 | 2 | 0.066 | 29 | 12 | -0.004 | 29 | 22 | 0.027 |
| 30 | 2 | 0.013 | 30 | 12 | -0.006 | 30 | 22 | -0.089 |
| 1 | 3 | 0.050 | 1 | 13 | -0.039 | 1 | 23 | -0.035 |
| 2 | 3 | -0.006 | 2 | 13 | -0.037 | 2 | 23 | 0.001 |
| 3 | 3 | -5.136 | 3 | 13 | -0.096 | 3 | 23 | -0.004 |
| 4 | 3 | -0.293 | 4 | 13 | -0.040 | 4 | 23 | -0.010 |
| 5 | 3 | 0.187 | 5 | 13 | -0.053 | 5 | 23 | 0.009 |
| 6 | 3 | -0.001 | 6 | 13 | -0.033 | 6 | 23 | 0.045 |
| 7 | 3 | -0.137 | 7 | 13 | 0.130 | 7 | 23 | -0.014 |
| 8 | 3 | 0.038 | 8 | 13 | 0.044 | 8 | 23 | -0.042 |
| 9 | 3 | -0.063 | 9 | 13 | -0.450 | 9 | 23 | -0.005 |
| 10 | 3 | -0.039 | 10 | 13 | -0.146 | 10 | 23 | -0.001 |
| 11 | 3 | -0.086 | 11 | 13 | -0.109 | 11 | 23 | 0.016 |
| 12 | 3 | 0.042 | 12 | 13 | 0.001 | 12 | 23 | 0.068 |
| 13 | 3 | -0.096 | 13 | 13 | -5.181 | 13 | 23 | 0.003 |
| 14 | 3 | 0.085 | 14 | 13 | 0.155 | 14 | 23 | 0.048 |
| 15 | 3 | -0.430 | 15 | 13 | 0.168 | 15 | 23 | -0.002 |
| 16 | 3 | -0.016 | 16 | 13 | 0.023 | 16 | 23 | 0.086 |
| 17 | 3 | 0.003 | 17 | 13 | 0.015 | 17 | 23 | -0.097 |
| 18 | 3 | 0.068 | 18 | 13 | -0.035 | 18 | 23 | 0.042 |
| 19 | 3 | -0.047 | 19 | 13 | 0.037 | 19 | 23 | -0.085 |
| 20 | 3 | -0.002 | 20 | 13 | -0.004 | 20 | 23 | -0.431 |
| 21 | 3 | 0.034 | 21 | 13 | 0.033 | 21 | 23 | -0.048 |
| 22 | 3 | -0.001 | 22 | 13 | -0.070 | 22 | 23 | 0.005 |
| 23 | 3 | -0.004 | 23 | 13 | 0.003 | 23 | 23 | -5.140 |
| 24 | 3 | 0.010 | 24 | 13 | -0.003 | 24 | 23 | 0.291 |
| 25 | 3 | 0.009 | 25 | 13 | 0.044 | 25 | 23 | 0.188 |
| 26 | 3 | -0.045 | 26 | 13 | -0.033 | 26 | 23 | 0.001 |
| 27 | 3 | 0.014 | 27 | 13 | 0.005 | 27 | 23 | 0.136 |
| 28 | 3 | -0.042 | 28 | 13 | -0.002 | 28 | 23 | 0.038 |
| 29 | 3 | 0.005 | 29 | 13 | 0.010 | 29 | 23 | 0.063 |
| 30 | 3 | -0.001 | 30 | 13 | -0.015 | 30 | 23 | -0.040 |
| 1 | 4 | -0.052 | 1 | 14 | -0.035 | 1 | 24 | 0.005 |
| 2 | 4 | 0.025 | 2 | 14 | -0.004 | 2 | 24 | -0.004 |
| 3 | 4 | -0.293 | 3 | 14 | 0.085 | 3 | 24 | 0.010 |
| 4 | 4 | -3.844 | 4 | 14 | 0.244 | 4 | 24 | -0.002 |
| 5 | 4 | -0.177 | 5 | 14 | -0.022 | 5 | 24 | 0.002 |
| 6 | 4 | -0.048 | 6 | 14 | -0.028 | 6 | 24 | 0.024 |
| 7 | 4 | 0.426 | 7 | 14 | 0.026 | 7 | 24 | 0.006 |
| 8 | 4 | -0.014 | 8 | 14 | -0.031 | 8 | 24 | -0.064 |
| 9 | 4 | 0.277 | 9 | 14 | 0.221 | 9 | 24 | -0.009 |
| 10 | 4 | -0.344 | 10 | 14 | 0.065 | 10 | 24 | -0.001 |
| 11 | 4 | -0.173 | 11 | 14 | 0.043 | 11 | 24 | -0.011 |
| 12 | 4 | 0.026 | 12 | 14 | -0.670 | 12 | 24 | -0.011 |
| 13 | 4 | -0.040 | 13 | 14 | 0.155 | 13 | 24 | -0.003 |
| 14 | 4 | 0.244 | 14 | 14 | -8.166 | 14 | 24 | -0.011 |
| 15 | 4 | -0.295 | 15 | 14 | 0.171 | 15 | 24 | -0.001 |
| 16 | 4 | -0.011 | 16 | 14 | 0.004 | 16 | 24 | -0.173 |
| 17 | 4 | 0.003 | 17 | 14 | -0.037 | 17 | 24 | 0.040 |
| 18 | 4 | 0.011 | 18 | 14 | 0.021 | 18 | 24 | -0.026 |
| 19 | 4 | -0.011 | 19 | 14 | 0.008 | 19 | 24 | 0.243 |
| 20 | 4 | 0.001 | 20 | 14 | -0.023 | 20 | 24 | 0.294 |
| 21 | 4 | 0.005 | 21 | 14 | -0.004 | 21 | 24 | -0.051 |
| 22 | 4 | -0.004 | 22 | 14 | -0.001 | 22 | 24 | 0.025 |
| 23 | 4 | -0.010 | 23 | 14 | 0.048 | 23 | 24 | 0.291 |
| 24 | 4 | -0.002 | 24 | 14 | -0.011 | 24 | 24 | -3.846 |
| 25 | 4 | -0.002 | 25 | 14 | 0.003 | 25 | 24 | 0.178 |
| 26 | 4 | 0.024 | 26 | 14 | -0.001 | 26 | 24 | -0.050 |
| 27 | 4 | 0.006 | 27 | 14 | -0.020 | 27 | 24 | 0.425 |
| 28 | 4 | 0.064 | 28 | 14 | -0.001 | 28 | 24 | 0.014 |
| 29 | 4 | -0.009 | 29 | 14 | 0.001 | 29 | 24 | 0.277 |
| 30 | 4 | 0.001 | 30 | 14 | 0.001 | 30 | 24 | 0.344 |
| 1 | 5 | -0.077 | 1 | 15 | 0.175 | 1 | 25 | -0.001 |
| 2 | 5 | -0.674 | 2 | 15 | -0.016 | 2 | 25 | -0.002 |
| 3 | 5 | 0.187 | 3 | 15 | -0.430 | 3 | 25 | 0.009 |
| 4 | 5 | -0.177 | 4 | 15 | -0.295 | 4 | 25 | -0.002 |
| 5 | 5 | -8.180 | 5 | 15 | 0.287 | 5 | 25 | 0.001 |
| 6 | 5 | -0.003 | 6 | 15 | 0.108 | 6 | 25 | 0.009 |
| 7 | 5 | 0.113 | 7 | 15 | -0.045 | 7 | 25 | 0.030 |
| 8 | 5 | 0.001 | 8 | 15 | -0.021 | 8 | 25 | 0.015 |
| 9 | 5 | 0.285 | 9 | 15 | 0.248 | 9 | 25 | -0.016 |
| 10 | 5 | -0.057 | 10 | 15 | 0.266 | 10 | 25 | 0.000 |
| 11 | 5 | 0.066 | 11 | 15 | -0.152 | 11 | 25 | -0.004 |
| 12 | 5 | 0.012 | 12 | 15 | -0.024 | 12 | 25 | -0.008 |
| 13 | 5 | -0.053 | 13 | 15 | 0.168 | 13 | 25 | 0.044 |
| 14 | 5 | -0.022 | 14 | 15 | 0.171 | 14 | 25 | 0.003 |
| 15 | 5 | 0.287 | 15 | 15 | -3.806 | 15 | 25 | -0.004 |
| 16 | 5 | 0.004 | 16 | 15 | 0.016 | 16 | 25 | -0.066 |
| 17 | 5 | 0.044 | 17 | 15 | -0.004 | 17 | 25 | -0.053 |
| 18 | 5 | -0.008 | 18 | 15 | -0.068 | 18 | 25 | 0.013 |
| 19 | 5 | -0.003 | 19 | 15 | 0.023 | 19 | 25 | 0.022 |
| 20 | 5 | -0.004 | 20 | 15 | -0.010 | 20 | 25 | 0.287 |
| 21 | 5 | 0.001 | 21 | 15 | -0.012 | 21 | 25 | 0.077 |
| 22 | 5 | 0.002 | 22 | 15 | 0.005 | 22 | 25 | 0.673 |
| 23 | 5 | 0.009 | 23 | 15 | -0.002 | 23 | 25 | 0.188 |
| 24 | 5 | 0.002 | 24 | 15 | -0.001 | 24 | 25 | 0.178 |
| 25 | 5 | 0.001 | 25 | 15 | -0.004 | 25 | 25 | -8.185 |
| 26 | 5 | -0.009 | 26 | 15 | 0.000 | 26 | 25 | 0.003 |
| 27 | 5 | -0.030 | 27 | 15 | -0.010 | 27 | 25 | -0.113 |
| 28 | 5 | 0.015 | 28 | 15 | 0.003 | 28 | 25 | 0.001 |
| 29 | 5 | 0.016 | 29 | 15 | 0.003 | 29 | 25 | -0.285 |
| 30 | 5 | 0.000 | 30 | 15 | 0.003 | 30 | 25 | -0.057 |
| 1 | 6 | -0.056 | 1 | 16 | -0.005 | 1 | 26 | 0.006 |
| 2 | 6 | -0.096 | 2 | 16 | -0.033 | 2 | 26 | 0.032 |
| 3 | 6 | -0.001 | 3 | 16 | -0.016 | 3 | 26 | -0.045 |
| 4 | 6 | -0.048 | 4 | 16 | -0.011 | 4 | 26 | 0.024 |
| 5 | 6 | -0.003 | 5 | 16 | 0.004 | 5 | 26 | -0.009 |
| 6 | 6 | -8.158 | 6 | 16 | 0.002 | 6 | 26 | -0.004 |
| 7 | 6 | 0.081 | 7 | 16 | -0.030 | 7 | 26 | 0.049 |
| 8 | 6 | 0.547 | 8 | 16 | 0.008 | 8 | 26 | -0.017 |
| 9 | 6 | -0.086 | 9 | 16 | -0.004 | 9 | 26 | 0.014 |
| 10 | 6 | 0.020 | 10 | 16 | -0.001 | 10 | 26 | 0.004 |
| 11 | 6 | -0.028 | 11 | 16 | -0.001 | 11 | 26 | 0.002 |
| 12 | 6 | 0.009 | 12 | 16 | -0.029 | 12 | 26 | -0.005 |
| 13 | 6 | -0.033 | 13 | 16 | 0.023 | 13 | 26 | -0.033 |
| 14 | 6 | -0.028 | 14 | 16 | 0.004 | 14 | 26 | -0.001 |
| 15 | 6 | 0.108 | 15 | 16 | 0.016 | 15 | 26 | 0.000 |
| 16 | 6 | 0.002 | 16 | 16 | -8.263 | 16 | 26 | -0.028 |
| 17 | 6 | 0.033 | 17 | 16 | 0.109 | 17 | 26 | 0.034 |
| 18 | 6 | 0.005 | 18 | 16 | -0.030 | 18 | 26 | -0.009 |
| 19 | 6 | -0.001 | 19 | 16 | 0.043 | 19 | 26 | -0.028 |
| 20 | 6 | 0.000 | 20 | 16 | 0.154 | 20 | 26 | -0.109 |
| 21 | 6 | 0.006 | 21 | 16 | -0.014 | 21 | 26 | -0.056 |
| 22 | 6 | 0.032 | 22 | 16 | -0.008 | 22 | 26 | -0.096 |
| 23 | 6 | 0.045 | 23 | 16 | 0.086 | 23 | 26 | 0.001 |
| 24 | 6 | 0.024 | 24 | 16 | -0.173 | 24 | 26 | -0.050 |
| 25 | 6 | 0.009 | 25 | 16 | -0.066 | 25 | 26 | 0.003 |
| 26 | 6 | -0.004 | 26 | 16 | -0.028 | 26 | 26 | -8.156 |
| 27 | 6 | 0.048 | 27 | 16 | -0.036 | 27 | 26 | 0.081 |
| 28 | 6 | 0.017 | 28 | 16 | 0.127 | 28 | 26 | -0.549 |
| 29 | 6 | 0.014 | 29 | 16 | -0.263 | 29 | 26 | -0.085 |
| 30 | 6 | -0.004 | 30 | 16 | 0.030 | 30 | 26 | -0.020 |
| 1 | 7 | 0.061 | 1 | 17 | -0.032 | 1 | 27 | 0.036 |
| 2 | 7 | 0.045 | 2 | 17 | 0.070 | 2 | 27 | -0.041 |
| 3 | 7 | -0.137 | 3 | 17 | 0.003 | 3 | 27 | 0.014 |
| 4 | 7 | 0.426 | 4 | 17 | 0.003 | 4 | 27 | 0.006 |
| 5 | 7 | 0.113 | 5 | 17 | 0.044 | 5 | 27 | -0.030 |
| 6 | 7 | 0.081 | 6 | 17 | 0.033 | 6 | 27 | 0.048 |
| 7 | 7 | -5.148 | 7 | 17 | -0.005 | 7 | 27 | 0.003 |
| 8 | 7 | 0.004 | 8 | 17 | -0.002 | 8 | 27 | -0.069 |
| 9 | 7 | -0.263 | 9 | 17 | -0.010 | 9 | 27 | 0.003 |
| 10 | 7 | 0.169 | 10 | 17 | -0.015 | 10 | 27 | 0.011 |
| 11 | 7 | -0.037 | 11 | 17 | -0.023 | 11 | 27 | -0.030 |
| 12 | 7 | 0.037 | 12 | 17 | -0.035 | 12 | 27 | -0.001 |
| 13 | 7 | 0.130 | 13 | 17 | 0.015 | 13 | 27 | 0.005 |
| 14 | 7 | 0.026 | 14 | 17 | -0.037 | 14 | 27 | -0.020 |
| 15 | 7 | -0.045 | 15 | 17 | -0.004 | 15 | 27 | -0.010 |
| 16 | 7 | -0.030 | 16 | 17 | 0.109 | 16 | 27 | -0.036 |
| 17 | 7 | -0.005 | 17 | 17 | -5.179 | 17 | 27 | -0.131 |
| 18 | 7 | 0.001 | 18 | 17 | 0.001 | 18 | 27 | -0.037 |
| 19 | 7 | -0.020 | 19 | 17 | -0.156 | 19 | 27 | 0.026 |
| 20 | 7 | 0.010 | 20 | 17 | 0.168 | 20 | 27 | 0.046 |
| 21 | 7 | 0.036 | 21 | 17 | 0.039 | 21 | 27 | 0.060 |
| 22 | 7 | -0.040 | 22 | 17 | 0.037 | 22 | 27 | 0.045 |
| 23 | 7 | -0.014 | 23 | 17 | -0.097 | 23 | 27 | 0.136 |
| 24 | 7 | 0.006 | 24 | 17 | 0.040 | 24 | 27 | 0.425 |
| 25 | 7 | 0.030 | 25 | 17 | -0.053 | 25 | 27 | -0.113 |
| 26 | 7 | 0.049 | 26 | 17 | 0.034 | 26 | 27 | 0.081 |
| 27 | 7 | 0.003 | 27 | 17 | -0.131 | 27 | 27 | -5.147 |
| 28 | 7 | 0.069 | 28 | 17 | 0.044 | 28 | 27 | -0.004 |
| 29 | 7 | 0.003 | 29 | 17 | 0.450 | 29 | 27 | -0.265 |
| 30 | 7 | -0.011 | 30 | 17 | -0.145 | 30 | 27 | -0.168 |
| 1 | 8 | -0.010 | 1 | 18 | -0.032 | 1 | 28 | -0.035 |
| 2 | 8 | 0.014 | 2 | 18 | 0.043 | 2 | 28 | 0.044 |
| 3 | 8 | 0.038 | 3 | 18 | 0.068 | 3 | 28 | -0.042 |
| 4 | 8 | -0.014 | 4 | 18 | 0.011 | 4 | 28 | 0.064 |
| 5 | 8 | 0.001 | 5 | 18 | -0.008 | 5 | 28 | 0.015 |
| 6 | 8 | 0.547 | 6 | 18 | 0.005 | 6 | 28 | 0.017 |
| 7 | 8 | 0.004 | 7 | 18 | 0.001 | 7 | 28 | 0.069 |
| 8 | 8 | -3.501 | 8 | 18 | -0.002 | 8 | 28 | -0.043 |
| 9 | 8 | -0.023 | 9 | 18 | 0.004 | 9 | 28 | 0.009 |
| 10 | 8 | 0.372 | 10 | 18 | -0.006 | 10 | 28 | -0.028 |
| 11 | 8 | -0.127 | 11 | 18 | 0.029 | 11 | 28 | -0.008 |
| 12 | 8 | -0.013 | 12 | 18 | 0.043 | 12 | 28 | -0.002 |
| 13 | 8 | 0.044 | 13 | 18 | -0.035 | 13 | 28 | -0.002 |
| 14 | 8 | -0.031 | 14 | 18 | 0.021 | 14 | 28 | -0.001 |
| 15 | 8 | -0.021 | 15 | 18 | -0.068 | 15 | 28 | 0.003 |
| 16 | 8 | 0.008 | 16 | 18 | -0.030 | 16 | 28 | 0.127 |
| 17 | 8 | -0.002 | 17 | 18 | 0.001 | 17 | 28 | 0.044 |
| 18 | 8 | -0.002 | 18 | 18 | -3.509 | 18 | 28 | -0.013 |
| 19 | 8 | 0.001 | 19 | 18 | 0.670 | 19 | 28 | 0.032 |
| 20 | 8 | 0.003 | 20 | 18 | -0.025 | 20 | 28 | -0.022 |
| 21 | 8 | 0.035 | 21 | 18 | -0.130 | 21 | 28 | 0.010 |
| 22 | 8 | -0.044 | 22 | 18 | 0.013 | 22 | 28 | -0.014 |
| 23 | 8 | -0.042 | 23 | 18 | 0.042 | 23 | 28 | 0.038 |
| 24 | 8 | -0.064 | 24 | 18 | -0.026 | 24 | 28 | 0.014 |
| 25 | 8 | 0.015 | 25 | 18 | 0.013 | 25 | 28 | 0.001 |
| 26 | 8 | -0.017 | 26 | 18 | -0.009 | 26 | 28 | -0.549 |
| 27 | 8 | -0.069 | 27 | 18 | -0.037 | 27 | 28 | -0.004 |
| 28 | 8 | -0.043 | 28 | 18 | -0.013 | 28 | 28 | -3.499 |
| 29 | 8 | -0.009 | 29 | 18 | 0.014 | 29 | 28 | 0.023 |
| 30 | 8 | -0.028 | 30 | 18 | -0.005 | 30 | 28 | 0.370 |
| 1 | 9 | 0.088 | 1 | 19 | -0.004 | 1 | 29 | -0.020 |
| 2 | 9 | 0.026 | 2 | 19 | -0.001 | 2 | 29 | 0.066 |
| 3 | 9 | -0.063 | 3 | 19 | -0.047 | 3 | 29 | 0.005 |
| 4 | 9 | 0.277 | 4 | 19 | -0.011 | 4 | 29 | -0.009 |
| 5 | 9 | 0.285 | 5 | 19 | -0.003 | 5 | 29 | 0.016 |
| 6 | 9 | -0.086 | 6 | 19 | -0.001 | 6 | 29 | 0.014 |
| 7 | 9 | -0.263 | 7 | 19 | -0.020 | 7 | 29 | 0.003 |
| 8 | 9 | -0.023 | 8 | 19 | 0.001 | 8 | 29 | -0.009 |
| 9 | 9 | -3.833 | 9 | 19 | 0.001 | 9 | 29 | -0.001 |
| 10 | 9 | -0.184 | 10 | 19 | -0.001 | 10 | 29 | -0.004 |
| 11 | 9 | -0.263 | 11 | 19 | 0.004 | 11 | 29 | -0.004 |
| 12 | 9 | -0.014 | 12 | 19 | -0.021 | 12 | 29 | -0.004 |
| 13 | 9 | -0.450 | 13 | 19 | 0.037 | 13 | 29 | 0.010 |
| 14 | 9 | 0.221 | 14 | 19 | 0.008 | 14 | 29 | 0.001 |
| 15 | 9 | 0.248 | 15 | 19 | 0.023 | 15 | 29 | 0.003 |
| 16 | 9 | -0.004 | 16 | 19 | 0.043 | 16 | 29 | -0.263 |
| 17 | 9 | -0.010 | 17 | 19 | -0.156 | 17 | 29 | 0.450 |
| 18 | 9 | 0.004 | 18 | 19 | 0.670 | 18 | 29 | 0.014 |
| 19 | 9 | 0.001 | 19 | 19 | -8.165 | 19 | 29 | 0.220 |
| 20 | 9 | -0.003 | 20 | 19 | -0.170 | 20 | 29 | -0.248 |
| 21 | 9 | -0.020 | 21 | 19 | -0.034 | 21 | 29 | 0.087 |
| 22 | 9 | 0.066 | 22 | 19 | -0.004 | 22 | 29 | 0.027 |
| 23 | 9 | -0.005 | 23 | 19 | -0.085 | 23 | 29 | 0.063 |
| 24 | 9 | -0.009 | 24 | 19 | 0.243 | 24 | 29 | 0.277 |
| 25 | 9 | -0.016 | 25 | 19 | 0.022 | 25 | 29 | -0.285 |
| 26 | 9 | 0.014 | 26 | 19 | -0.028 | 26 | 29 | -0.085 |
| 27 | 9 | 0.003 | 27 | 19 | 0.026 | 27 | 29 | -0.265 |
| 28 | 9 | 0.009 | 28 | 19 | 0.032 | 28 | 29 | 0.023 |
| 29 | 9 | -0.001 | 29 | 19 | 0.220 | 29 | 29 | -3.832 |
| 30 | 9 | 0.004 | 30 | 19 | -0.065 | 30 | 29 | 0.183 |
| 1 | 10 | -0.031 | 1 | 20 | 0.012 | 1 | 30 | -0.003 |
| 2 | 10 | 0.089 | 2 | 20 | -0.005 | 2 | 30 | 0.013 |
| 3 | 10 | -0.039 | 3 | 20 | -0.002 | 3 | 30 | -0.001 |
| 4 | 10 | -0.344 | 4 | 20 | 0.001 | 4 | 30 | 0.001 |
| 5 | 10 | -0.057 | 5 | 20 | -0.004 | 5 | 30 | 0.000 |
| 6 | 10 | 0.020 | 6 | 20 | 0.000 | 6 | 30 | -0.004 |
| 7 | 10 | 0.169 | 7 | 20 | 0.010 | 7 | 30 | -0.011 |
| 8 | 10 | 0.372 | 8 | 20 | 0.003 | 8 | 30 | -0.028 |
| 9 | 10 | -0.184 | 9 | 20 | -0.003 | 9 | 30 | 0.004 |
| 10 | 10 | -8.290 | 10 | 20 | 0.003 | 10 | 30 | -0.002 |
| 11 | 10 | -0.030 | 11 | 20 | -0.016 | 11 | 30 | 0.001 |
| 12 | 10 | -0.005 | 12 | 20 | -0.068 | 12 | 30 | -0.006 |
| 13 | 10 | -0.146 | 13 | 20 | -0.004 | 13 | 30 | -0.015 |
| 14 | 10 | 0.065 | 14 | 20 | -0.023 | 14 | 30 | 0.001 |
| 15 | 10 | 0.266 | 15 | 20 | -0.010 | 15 | 30 | 0.003 |
| 16 | 10 | -0.001 | 16 | 20 | 0.154 | 16 | 30 | 0.030 |
| 17 | 10 | -0.015 | 17 | 20 | 0.168 | 17 | 30 | -0.145 |
| 18 | 10 | -0.006 | 18 | 20 | -0.025 | 18 | 30 | -0.005 |
| 19 | 10 | -0.001 | 19 | 20 | -0.170 | 19 | 30 | -0.065 |
| 20 | 10 | 0.003 | 20 | 20 | -3.805 | 20 | 30 | 0.265 |
| 21 | 10 | 0.003 | 21 | 20 | -0.173 | 21 | 30 | 0.031 |
| 22 | 10 | -0.013 | 22 | 20 | 0.016 | 22 | 30 | -0.089 |
| 23 | 10 | -0.001 | 23 | 20 | -0.431 | 23 | 30 | -0.040 |
| 24 | 10 | -0.001 | 24 | 20 | 0.294 | 24 | 30 | 0.344 |
| 25 | 10 | 0.000 | 25 | 20 | 0.287 | 25 | 30 | -0.057 |
| 26 | 10 | 0.004 | 26 | 20 | -0.109 | 26 | 30 | -0.020 |
| 27 | 10 | 0.011 | 27 | 20 | 0.046 | 27 | 30 | -0.168 |
| 28 | 10 | -0.028 | 28 | 20 | -0.022 | 28 | 30 | 0.370 |
| 29 | 10 | -0.004 | 29 | 20 | -0.248 | 29 | 30 | 0.183 |
| 30 | 10 | -0.002 | 30 | 20 | 0.265 | 30 | 30 | -8.289 |

**Tight-binding Hamiltonian for ([$Mn_3$]$_2$-(ppmd)$_3$)$^{2+}$**

| | | | | | | | | |
|---|---|---|---|---|---|---|---|---|
| 1 | 1 | -8.085 | 1 | 11 | -0.019 | 1 | 21 | 0.001 |
| 2 | 1 | -0.022 | 2 | 11 | 0.012 | 2 | 21 | 0.000 |
| 3 | 1 | -0.552 | 3 | 11 | 0.082 | 3 | 21 | -0.006 |
| 4 | 1 | -0.043 | 4 | 11 | 0.028 | 4 | 21 | 0.000 |
| 5 | 1 | 0.138 | 5 | 11 | -0.112 | 5 | 21 | 0.001 |
| 6 | 1 | -0.032 | 6 | 11 | -0.039 | 6 | 21 | 0.000 |
| 7 | 1 | -0.041 | 7 | 11 | -0.060 | 7 | 21 | -0.002 |
| 8 | 1 | -0.025 | 8 | 11 | -0.010 | 8 | 21 | 0.000 |
| 9 | 1 | 0.150 | 9 | 11 | 0.019 | 9 | 21 | 0.001 |
| 10 | 1 | -0.058 | 10 | 11 | 0.026 | 10 | 21 | 0.000 |
| 11 | 1 | -0.019 | 11 | 11 | -8.069 | 11 | 21 | 0.000 |
| 12 | 1 | 0.007 | 12 | 11 | -0.339 | 12 | 21 | 0.000 |
| 13 | 1 | 0.012 | 13 | 11 | 0.086 | 13 | 21 | 0.004 |
| 14 | 1 | 0.168 | 14 | 11 | -0.160 | 14 | 21 | 0.000 |
| 15 | 1 | 0.016 | 15 | 11 | 0.015 | 15 | 21 | 0.000 |
| 16 | 1 | 0.000 | 16 | 11 | 0.000 | 16 | 21 | -0.019 |
| 17 | 1 | 0.000 | 17 | 11 | 0.000 | 17 | 21 | 0.007 |
| 18 | 1 | -0.004 | 18 | 11 | 0.003 | 18 | 21 | -0.012 |
| 19 | 1 | 0.000 | 19 | 11 | -0.002 | 19 | 21 | 0.168 |
| 20 | 1 | 0.000 | 20 | 11 | 0.000 | 20 | 21 | -0.016 |
| 21 | 1 | 0.001 | 21 | 11 | 0.000 | 21 | 21 | -8.083 |
| 22 | 1 | 0.006 | 22 | 11 | -0.002 | 22 | 21 | 0.552 |
| 23 | 1 | 0.000 | 23 | 11 | 0.006 | 23 | 21 | 0.022 |
| 24 | 1 | 0.000 | 24 | 11 | 0.000 | 24 | 21 | -0.043 |
| 25 | 1 | -0.001 | 25 | 11 | 0.001 | 25 | 21 | -0.138 |
| 26 | 1 | 0.000 | 26 | 11 | 0.001 | 26 | 21 | -0.032 |
| 27 | 1 | -0.002 | 27 | 11 | 0.000 | 27 | 21 | -0.041 |
| 28 | 1 | 0.000 | 28 | 11 | -0.007 | 28 | 21 | 0.026 |
| 29 | 1 | 0.001 | 29 | 11 | 0.001 | 29 | 21 | 0.150 |
| 30 | 1 | 0.000 | 30 | 11 | 0.001 | 30 | 21 | 0.057 |
| 1 | 2 | -0.022 | 1 | 12 | 0.007 | 1 | 22 | 0.006 |
| 2 | 2 | -5.034 | 2 | 12 | 0.041 | 2 | 22 | 0.000 |
| 3 | 2 | 0.000 | 3 | 12 | 0.011 | 3 | 22 | 0.035 |
| 4 | 2 | 0.099 | 4 | 12 | -0.006 | 4 | 22 | -0.003 |
| 5 | 2 | -0.256 | 5 | 12 | -0.029 | 5 | 22 | 0.002 |
| 6 | 2 | 0.031 | 6 | 12 | 0.064 | 6 | 22 | 0.000 |
| 7 | 2 | -0.137 | 7 | 12 | 0.158 | 7 | 22 | 0.005 |
| 8 | 2 | 0.156 | 8 | 12 | -0.011 | 8 | 22 | 0.002 |
| 9 | 2 | -0.386 | 9 | 12 | -0.060 | 9 | 22 | -0.004 |
| 10 | 2 | 0.116 | 10 | 12 | 0.072 | 10 | 22 | 0.000 |
| 11 | 2 | 0.012 | 11 | 12 | -0.339 | 11 | 22 | -0.002 |
| 12 | 2 | 0.041 | 12 | 12 | -3.271 | 12 | 22 | 0.016 |
| 13 | 2 | -0.117 | 13 | 12 | -0.002 | 13 | 22 | -0.018 |
| 14 | 2 | -0.057 | 14 | 12 | 0.057 | 14 | 22 | -0.005 |
| 15 | 2 | 0.063 | 15 | 12 | -0.449 | 15 | 22 | 0.002 |
| 16 | 2 | -0.006 | 16 | 12 | 0.000 | 16 | 22 | -0.082 |
| 17 | 2 | -0.019 | 17 | 12 | 0.002 | 17 | 22 | -0.011 |
| 18 | 2 | 0.008 | 18 | 12 | 0.005 | 18 | 22 | 0.153 |
| 19 | 2 | 0.006 | 19 | 12 | -0.004 | 19 | 22 | -0.062 |
| 20 | 2 | 0.000 | 20 | 12 | 0.000 | 20 | 22 | -0.054 |
| 21 | 2 | 0.000 | 21 | 12 | 0.000 | 21 | 22 | 0.552 |
| 22 | 2 | 0.000 | 22 | 12 | 0.016 | 22 | 22 | -3.266 |
| 23 | 2 | -0.003 | 23 | 12 | 0.019 | 23 | 22 | 0.000 |
| 24 | 2 | 0.000 | 24 | 12 | -0.003 | 24 | 22 | 0.111 |
| 25 | 2 | 0.001 | 25 | 12 | 0.004 | 25 | 22 | -0.054 |
| 26 | 2 | 0.003 | 26 | 12 | -0.007 | 26 | 22 | -0.010 |
| 27 | 2 | 0.000 | 27 | 12 | 0.000 | 27 | 22 | 0.041 |
| 28 | 2 | 0.005 | 28 | 12 | -0.035 | 28 | 22 | -0.011 |
| 29 | 2 | 0.000 | 29 | 12 | -0.002 | 29 | 22 | 0.027 |
| 30 | 2 | -0.002 | 30 | 12 | 0.000 | 30 | 22 | 0.002 |
| 1 | 3 | -0.552 | 1 | 13 | 0.012 | 1 | 23 | 0.000 |
| 2 | 3 | 0.000 | 2 | 13 | -0.117 | 2 | 23 | -0.003 |
| 3 | 3 | -3.267 | 3 | 13 | 0.154 | 3 | 23 | 0.000 |
| 4 | 3 | -0.110 | 4 | 13 | 0.105 | 4 | 23 | 0.000 |
| 5 | 3 | -0.054 | 5 | 13 | 0.373 | 5 | 23 | 0.001 |
| 6 | 3 | 0.010 | 6 | 13 | -0.027 | 6 | 23 | -0.003 |
| 7 | 3 | -0.041 | 7 | 13 | 0.140 | 7 | 23 | 0.000 |
| 8 | 3 | -0.011 | 8 | 13 | 0.042 | 8 | 23 | 0.005 |
| 9 | 3 | -0.027 | 9 | 13 | -0.070 | 9 | 23 | 0.000 |
| 10 | 3 | 0.002 | 10 | 13 | 0.035 | 10 | 23 | -0.002 |
| 11 | 3 | 0.082 | 11 | 13 | 0.086 | 11 | 23 | 0.006 |
| 12 | 3 | 0.011 | 12 | 13 | -0.002 | 12 | 23 | 0.019 |
| 13 | 3 | 0.154 | 13 | 13 | -4.998 | 13 | 23 | 0.008 |
| 14 | 3 | 0.062 | 14 | 13 | -0.344 | 14 | 23 | -0.006 |
| 15 | 3 | -0.054 | 15 | 13 | -0.044 | 15 | 23 | 0.000 |
| 16 | 3 | 0.002 | 16 | 13 | -0.003 | 16 | 23 | -0.012 |
| 17 | 3 | -0.016 | 17 | 13 | -0.005 | 17 | 23 | -0.041 |
| 18 | 3 | -0.018 | 18 | 13 | 0.000 | 18 | 23 | -0.117 |
| 19 | 3 | 0.005 | 19 | 13 | 0.000 | 19 | 23 | 0.057 |
| 20 | 3 | 0.002 | 20 | 13 | -0.001 | 20 | 23 | 0.063 |
| 21 | 3 | -0.006 | 21 | 13 | 0.004 | 21 | 23 | 0.022 |
| 22 | 3 | 0.035 | 22 | 13 | -0.018 | 22 | 23 | 0.000 |
| 23 | 3 | 0.000 | 23 | 13 | 0.008 | 23 | 23 | -5.033 |
| 24 | 3 | 0.003 | 24 | 13 | -0.004 | 24 | 23 | -0.099 |
| 25 | 3 | 0.002 | 25 | 13 | -0.006 | 25 | 23 | -0.257 |
| 26 | 3 | 0.000 | 26 | 13 | 0.000 | 26 | 23 | -0.031 |
| 27 | 3 | -0.005 | 27 | 13 | 0.003 | 27 | 23 | 0.136 |
| 28 | 3 | 0.002 | 28 | 13 | 0.000 | 28 | 23 | 0.156 |
| 29 | 3 | 0.004 | 29 | 13 | 0.001 | 29 | 23 | 0.386 |
| 30 | 3 | 0.000 | 30 | 13 | 0.000 | 30 | 23 | 0.116 |
| 1 | 4 | -0.043 | 1 | 14 | 0.168 | 1 | 24 | 0.000 |
| 2 | 4 | 0.099 | 2 | 14 | -0.057 | 2 | 24 | 0.000 |
| 3 | 4 | -0.110 | 3 | 14 | 0.062 | 3 | 24 | 0.003 |
| 4 | 4 | -8.139 | 4 | 14 | 0.226 | 4 | 24 | -0.001 |
| 5 | 4 | -0.118 | 5 | 14 | 0.271 | 5 | 24 | -0.001 |
| 6 | 4 | 0.005 | 6 | 14 | 0.012 | 6 | 24 | 0.000 |
| 7 | 4 | -0.027 | 7 | 14 | -0.371 | 7 | 24 | 0.003 |
| 8 | 4 | 0.093 | 8 | 14 | -0.026 | 8 | 24 | 0.000 |
| 9 | 4 | 0.349 | 9 | 14 | 0.260 | 9 | 24 | -0.001 |
| 10 | 4 | -0.041 | 10 | 14 | 0.381 | 10 | 24 | 0.000 |
| 11 | 4 | 0.028 | 11 | 14 | -0.160 | 11 | 24 | 0.000 |
| 12 | 4 | -0.006 | 12 | 14 | 0.057 | 12 | 24 | -0.003 |
| 13 | 4 | 0.105 | 13 | 14 | -0.344 | 13 | 24 | -0.004 |
| 14 | 4 | 0.226 | 14 | 14 | -3.586 | 14 | 24 | -0.002 |
| 15 | 4 | 0.070 | 15 | 14 | 0.000 | 15 | 24 | 0.000 |
| 16 | 4 | 0.000 | 16 | 14 | -0.002 | 16 | 24 | 0.028 |
| 17 | 4 | -0.003 | 17 | 14 | -0.004 | 17 | 24 | -0.006 |
| 18 | 4 | 0.004 | 18 | 14 | 0.000 | 18 | 24 | -0.105 |
| 19 | 4 | -0.002 | 19 | 14 | 0.001 | 19 | 24 | 0.226 |
| 20 | 4 | 0.000 | 20 | 14 | 0.000 | 20 | 24 | -0.070 |
| 21 | 4 | 0.000 | 21 | 14 | 0.000 | 21 | 24 | -0.043 |
| 22 | 4 | -0.003 | 22 | 14 | -0.005 | 22 | 24 | 0.111 |
| 23 | 4 | 0.000 | 23 | 14 | -0.006 | 23 | 24 | -0.099 |
| 24 | 4 | -0.001 | 24 | 14 | -0.002 | 24 | 24 | -8.137 |
| 25 | 4 | 0.001 | 25 | 14 | -0.002 | 25 | 24 | 0.117 |
| 26 | 4 | 0.000 | 26 | 14 | -0.001 | 26 | 24 | 0.005 |
| 27 | 4 | 0.003 | 27 | 14 | -0.001 | 27 | 24 | -0.027 |
| 28 | 4 | 0.000 | 28 | 14 | 0.002 | 28 | 24 | -0.093 |
| 29 | 4 | -0.001 | 29 | 14 | -0.004 | 29 | 24 | 0.349 |
| 30 | 4 | 0.000 | 30 | 14 | 0.001 | 30 | 24 | 0.041 |
| 1 | 5 | 0.138 | 1 | 15 | 0.016 | 1 | 25 | -0.001 |
| 2 | 5 | -0.256 | 2 | 15 | 0.063 | 2 | 25 | 0.001 |
| 3 | 5 | -0.054 | 3 | 15 | -0.054 | 3 | 25 | 0.002 |
| 4 | 5 | -0.118 | 4 | 15 | 0.070 | 4 | 25 | 0.001 |
| 5 | 5 | -3.549 | 5 | 15 | 0.358 | 5 | 25 | -0.004 |
| 6 | 5 | -0.057 | 6 | 15 | 0.025 | 6 | 25 | -0.001 |
| 7 | 5 | -0.058 | 7 | 15 | 0.091 | 7 | 25 | 0.000 |
| 8 | 5 | 0.052 | 8 | 15 | -0.001 | 8 | 25 | 0.004 |
| 9 | 5 | -0.245 | 9 | 15 | -0.281 | 9 | 25 | -0.001 |
| 10 | 5 | 0.268 | 10 | 15 | -0.059 | 10 | 25 | -0.001 |
| 11 | 5 | -0.112 | 11 | 15 | 0.015 | 11 | 25 | 0.001 |
| 12 | 5 | -0.029 | 12 | 15 | -0.449 | 12 | 25 | 0.004 |
| 13 | 5 | 0.373 | 13 | 15 | -0.044 | 13 | 25 | -0.006 |
| 14 | 5 | 0.271 | 14 | 15 | 0.000 | 14 | 25 | -0.002 |
| 15 | 5 | 0.358 | 15 | 15 | -8.165 | 15 | 25 | -0.002 |
| 16 | 5 | -0.001 | 16 | 15 | 0.000 | 16 | 25 | 0.112 |
| 17 | 5 | -0.004 | 17 | 15 | 0.000 | 17 | 25 | 0.029 |
| 18 | 5 | -0.006 | 18 | 15 | -0.001 | 18 | 25 | 0.373 |
| 19 | 5 | 0.002 | 19 | 15 | 0.000 | 19 | 25 | -0.271 |
| 20 | 5 | -0.002 | 20 | 15 | 0.000 | 20 | 25 | 0.358 |
| 21 | 5 | 0.001 | 21 | 15 | 0.000 | 21 | 25 | -0.138 |
| 22 | 5 | 0.002 | 22 | 15 | 0.002 | 22 | 25 | -0.054 |
| 23 | 5 | 0.001 | 23 | 15 | 0.000 | 23 | 25 | -0.257 |
| 24 | 5 | -0.001 | 24 | 15 | 0.000 | 24 | 25 | 0.117 |
| 25 | 5 | -0.004 | 25 | 15 | -0.002 | 25 | 25 | -3.548 |
| 26 | 5 | 0.001 | 26 | 15 | 0.001 | 26 | 25 | 0.057 |
| 27 | 5 | 0.000 | 27 | 15 | 0.000 | 27 | 25 | 0.058 |
| 28 | 5 | 0.004 | 28 | 15 | -0.002 | 28 | 25 | 0.052 |
| 29 | 5 | 0.001 | 29 | 15 | 0.000 | 29 | 25 | 0.245 |
| 30 | 5 | -0.001 | 30 | 15 | 0.000 | 30 | 25 | 0.268 |
| 1 | 6 | -0.032 | 1 | 16 | 0.000 | 1 | 26 | 0.000 |
| 2 | 6 | 0.031 | 2 | 16 | -0.006 | 2 | 26 | 0.003 |
| 3 | 6 | 0.010 | 3 | 16 | 0.002 | 3 | 26 | 0.000 |
| 4 | 6 | 0.005 | 4 | 16 | 0.000 | 4 | 26 | 0.000 |
| 5 | 6 | -0.057 | 5 | 16 | -0.001 | 5 | 26 | 0.001 |
| 6 | 6 | -8.066 | 6 | 16 | 0.001 | 6 | 26 | 0.000 |
| 7 | 6 | 0.057 | 7 | 16 | 0.000 | 7 | 26 | 0.006 |
| 8 | 6 | 0.451 | 8 | 16 | 0.007 | 8 | 26 | -0.002 |
| 9 | 6 | 0.154 | 9 | 16 | 0.001 | 9 | 26 | -0.001 |
| 10 | 6 | 0.008 | 10 | 16 | -0.001 | 10 | 26 | 0.000 |
| 11 | 6 | -0.039 | 11 | 16 | 0.000 | 11 | 26 | 0.001 |
| 12 | 6 | 0.064 | 12 | 16 | 0.000 | 12 | 26 | -0.007 |
| 13 | 6 | -0.027 | 13 | 16 | -0.003 | 13 | 26 | 0.000 |
| 14 | 6 | 0.012 | 14 | 16 | -0.002 | 14 | 26 | -0.001 |
| 15 | 6 | 0.025 | 15 | 16 | 0.000 | 15 | 26 | 0.001 |
| 16 | 6 | 0.001 | 16 | 16 | -8.069 | 16 | 26 | -0.039 |
| 17 | 6 | -0.007 | 17 | 16 | -0.339 | 17 | 26 | 0.064 |
| 18 | 6 | 0.000 | 18 | 16 | -0.087 | 18 | 26 | 0.027 |
| 19 | 6 | -0.001 | 19 | 16 | -0.160 | 19 | 26 | 0.012 |
| 20 | 6 | -0.001 | 20 | 16 | -0.015 | 20 | 26 | -0.025 |
| 21 | 6 | 0.000 | 21 | 16 | -0.019 | 21 | 26 | -0.032 |
| 22 | 6 | 0.000 | 22 | 16 | -0.082 | 22 | 26 | -0.010 |
| 23 | 6 | -0.003 | 23 | 16 | -0.012 | 23 | 26 | -0.031 |
| 24 | 6 | 0.000 | 24 | 16 | 0.028 | 24 | 26 | 0.005 |
| 25 | 6 | -0.001 | 25 | 16 | 0.112 | 25 | 26 | 0.057 |
| 26 | 6 | 0.000 | 26 | 16 | -0.039 | 26 | 26 | -8.067 |
| 27 | 6 | 0.006 | 27 | 16 | -0.060 | 27 | 26 | 0.056 |
| 28 | 6 | 0.002 | 28 | 16 | 0.010 | 28 | 26 | -0.451 |
| 29 | 6 | -0.001 | 29 | 16 | 0.019 | 29 | 26 | 0.153 |
| 30 | 6 | 0.000 | 30 | 16 | -0.026 | 30 | 26 | -0.008 |
| 1 | 7 | -0.041 | 1 | 17 | 0.000 | 1 | 27 | -0.002 |
| 2 | 7 | -0.137 | 2 | 17 | -0.019 | 2 | 27 | 0.000 |
| 3 | 7 | -0.041 | 3 | 17 | -0.016 | 3 | 27 | -0.005 |
| 4 | 7 | -0.027 | 4 | 17 | -0.003 | 4 | 27 | 0.003 |
| 5 | 7 | -0.058 | 5 | 17 | -0.004 | 5 | 27 | 0.000 |
| 6 | 7 | 0.057 | 6 | 17 | -0.007 | 6 | 27 | 0.006 |
| 7 | 7 | -5.008 | 7 | 17 | 0.000 | 7 | 27 | 0.008 |
| 8 | 7 | 0.003 | 8 | 17 | 0.035 | 8 | 27 | -0.018 |
| 9 | 7 | 0.321 | 9 | 17 | -0.002 | 9 | 27 | 0.006 |
| 10 | 7 | 0.072 | 10 | 17 | 0.000 | 10 | 27 | 0.002 |
| 11 | 7 | -0.060 | 11 | 17 | 0.000 | 11 | 27 | 0.000 |
| 12 | 7 | 0.158 | 12 | 17 | 0.002 | 12 | 27 | 0.000 |
| 13 | 7 | 0.140 | 13 | 17 | -0.005 | 13 | 27 | 0.003 |
| 14 | 7 | -0.371 | 14 | 17 | -0.004 | 14 | 27 | -0.001 |
| 15 | 7 | 0.091 | 15 | 17 | 0.000 | 15 | 27 | 0.000 |
| 16 | 7 | 0.000 | 16 | 17 | -0.339 | 16 | 27 | -0.060 |
| 17 | 7 | 0.000 | 17 | 17 | -3.269 | 17 | 27 | 0.158 |
| 18 | 7 | -0.003 | 18 | 17 | 0.002 | 18 | 27 | -0.140 |
| 19 | 7 | -0.001 | 19 | 17 | 0.057 | 19 | 27 | -0.372 |
| 20 | 7 | 0.000 | 20 | 17 | 0.449 | 20 | 27 | -0.091 |
| 21 | 7 | -0.002 | 21 | 17 | 0.007 | 21 | 27 | -0.041 |
| 22 | 7 | 0.005 | 22 | 17 | -0.011 | 22 | 27 | 0.041 |
| 23 | 7 | 0.000 | 23 | 17 | -0.041 | 23 | 27 | 0.136 |
| 24 | 7 | 0.003 | 24 | 17 | -0.006 | 24 | 27 | -0.027 |
| 25 | 7 | 0.000 | 25 | 17 | 0.029 | 25 | 27 | 0.058 |
| 26 | 7 | 0.006 | 26 | 17 | 0.064 | 26 | 27 | 0.056 |
| 27 | 7 | 0.008 | 27 | 17 | 0.158 | 27 | 27 | -5.008 |
| 28 | 7 | 0.018 | 28 | 17 | 0.011 | 28 | 27 | -0.003 |
| 29 | 7 | 0.006 | 29 | 17 | -0.060 | 29 | 27 | 0.320 |
| 30 | 7 | -0.002 | 30 | 17 | -0.072 | 30 | 27 | -0.072 |
| 1 | 8 | -0.025 | 1 | 18 | -0.004 | 1 | 28 | 0.000 |
| 2 | 8 | 0.156 | 2 | 18 | 0.008 | 2 | 28 | 0.005 |
| 3 | 8 | -0.011 | 3 | 18 | -0.018 | 3 | 28 | 0.002 |
| 4 | 8 | 0.093 | 4 | 18 | 0.004 | 4 | 28 | 0.000 |
| 5 | 8 | 0.052 | 5 | 18 | -0.006 | 5 | 28 | 0.004 |
| 6 | 8 | 0.451 | 6 | 18 | 0.000 | 6 | 28 | 0.002 |
| 7 | 8 | 0.003 | 7 | 18 | -0.003 | 7 | 28 | 0.018 |
| 8 | 8 | -3.269 | 8 | 18 | 0.000 | 8 | 28 | 0.016 |
| 9 | 8 | 0.056 | 9 | 18 | -0.001 | 9 | 28 | -0.005 |
| 10 | 8 | -0.334 | 10 | 18 | 0.000 | 10 | 28 | 0.002 |
| 11 | 8 | -0.010 | 11 | 18 | 0.003 | 11 | 28 | -0.007 |
| 12 | 8 | -0.011 | 12 | 18 | 0.005 | 12 | 28 | -0.035 |
| 13 | 8 | 0.042 | 13 | 18 | 0.000 | 13 | 28 | 0.000 |
| 14 | 8 | -0.026 | 14 | 18 | 0.000 | 14 | 28 | 0.002 |
| 15 | 8 | -0.001 | 15 | 18 | -0.001 | 15 | 28 | -0.002 |
| 16 | 8 | 0.007 | 16 | 18 | -0.087 | 16 | 28 | 0.010 |
| 17 | 8 | 0.035 | 17 | 18 | 0.002 | 17 | 28 | 0.011 |
| 18 | 8 | 0.000 | 18 | 18 | -4.997 | 18 | 28 | 0.042 |
| 19 | 8 | -0.002 | 19 | 18 | 0.345 | 19 | 28 | 0.026 |
| 20 | 8 | -0.002 | 20 | 18 | -0.044 | 20 | 28 | -0.001 |
| 21 | 8 | 0.000 | 21 | 18 | -0.012 | 21 | 28 | 0.026 |
| 22 | 8 | 0.002 | 22 | 18 | 0.153 | 22 | 28 | -0.011 |
| 23 | 8 | 0.005 | 23 | 18 | -0.117 | 23 | 28 | 0.156 |
| 24 | 8 | 0.000 | 24 | 18 | -0.105 | 24 | 28 | -0.093 |
| 25 | 8 | 0.004 | 25 | 18 | 0.373 | 25 | 28 | 0.052 |
| 26 | 8 | -0.002 | 26 | 18 | 0.027 | 26 | 28 | -0.451 |
| 27 | 8 | -0.018 | 27 | 18 | -0.140 | 27 | 28 | -0.003 |
| 28 | 8 | 0.016 | 28 | 18 | 0.042 | 28 | 28 | -3.270 |
| 29 | 8 | 0.005 | 29 | 18 | 0.070 | 29 | 28 | -0.056 |
| 30 | 8 | 0.002 | 30 | 18 | 0.035 | 30 | 28 | -0.334 |
| 1 | 9 | 0.150 | 1 | 19 | 0.000 | 1 | 29 | 0.001 |
| 2 | 9 | -0.386 | 2 | 19 | 0.006 | 2 | 29 | 0.000 |
| 3 | 9 | -0.027 | 3 | 19 | 0.005 | 3 | 29 | 0.004 |
| 4 | 9 | 0.349 | 4 | 19 | -0.002 | 4 | 29 | -0.001 |
| 5 | 9 | -0.245 | 5 | 19 | 0.002 | 5 | 29 | 0.001 |
| 6 | 9 | 0.154 | 6 | 19 | -0.001 | 6 | 29 | -0.001 |
| 7 | 9 | 0.321 | 7 | 19 | -0.001 | 7 | 29 | 0.006 |
| 8 | 9 | 0.056 | 8 | 19 | -0.002 | 8 | 29 | 0.005 |
| 9 | 9 | -3.574 | 9 | 19 | -0.004 | 9 | 29 | -0.002 |
| 10 | 9 | 0.043 | 10 | 19 | -0.001 | 10 | 29 | -0.002 |
| 11 | 9 | 0.019 | 11 | 19 | -0.002 | 11 | 29 | 0.001 |
| 12 | 9 | -0.060 | 12 | 19 | -0.004 | 12 | 29 | -0.002 |
| 13 | 9 | -0.070 | 13 | 19 | 0.000 | 13 | 29 | 0.001 |
| 14 | 9 | 0.260 | 14 | 19 | 0.001 | 14 | 29 | -0.004 |
| 15 | 9 | -0.281 | 15 | 19 | 0.000 | 15 | 29 | 0.000 |
| 16 | 9 | 0.001 | 16 | 19 | -0.160 | 16 | 29 | 0.019 |
| 17 | 9 | -0.002 | 17 | 19 | 0.057 | 17 | 29 | -0.060 |
| 18 | 9 | -0.001 | 18 | 19 | 0.345 | 18 | 29 | 0.070 |
| 19 | 9 | -0.004 | 19 | 19 | -3.585 | 19 | 29 | 0.259 |
| 20 | 9 | 0.000 | 20 | 19 | 0.001 | 20 | 29 | 0.281 |
| 21 | 9 | 0.001 | 21 | 19 | 0.168 | 21 | 29 | 0.150 |
| 22 | 9 | -0.004 | 22 | 19 | -0.062 | 22 | 29 | 0.027 |
| 23 | 9 | 0.000 | 23 | 19 | 0.057 | 23 | 29 | 0.386 |
| 24 | 9 | -0.001 | 24 | 19 | 0.226 | 24 | 29 | 0.349 |
| 25 | 9 | -0.001 | 25 | 19 | -0.271 | 25 | 29 | 0.245 |
| 26 | 9 | -0.001 | 26 | 19 | 0.012 | 26 | 29 | 0.153 |
| 27 | 9 | 0.006 | 27 | 19 | -0.372 | 27 | 29 | 0.320 |
| 28 | 9 | -0.005 | 28 | 19 | 0.026 | 28 | 29 | -0.056 |
| 29 | 9 | -0.002 | 29 | 19 | 0.259 | 29 | 29 | -3.574 |
| 30 | 9 | 0.002 | 30 | 19 | -0.381 | 30 | 29 | -0.043 |
| 1 | 10 | -0.058 | 1 | 20 | 0.000 | 1 | 30 | 0.000 |
| 2 | 10 | 0.116 | 2 | 20 | 0.000 | 2 | 30 | -0.002 |
| 3 | 10 | 0.002 | 3 | 20 | 0.002 | 3 | 30 | 0.000 |
| 4 | 10 | -0.041 | 4 | 20 | 0.000 | 4 | 30 | 0.000 |
| 5 | 10 | 0.268 | 5 | 20 | -0.002 | 5 | 30 | -0.001 |
| 6 | 10 | 0.008 | 6 | 20 | -0.001 | 6 | 30 | 0.000 |
| 7 | 10 | 0.072 | 7 | 20 | 0.000 | 7 | 30 | -0.002 |
| 8 | 10 | -0.334 | 8 | 20 | -0.002 | 8 | 30 | 0.002 |
| 9 | 10 | 0.043 | 9 | 20 | 0.000 | 9 | 30 | 0.002 |
| 10 | 10 | -8.166 | 10 | 20 | 0.000 | 10 | 30 | 0.000 |
| 11 | 10 | 0.026 | 11 | 20 | 0.000 | 11 | 30 | 0.001 |
| 12 | 10 | 0.072 | 12 | 20 | 0.000 | 12 | 30 | 0.000 |
| 13 | 10 | 0.035 | 13 | 20 | -0.001 | 13 | 30 | 0.000 |
| 14 | 10 | 0.381 | 14 | 20 | 0.000 | 14 | 30 | 0.001 |
| 15 | 10 | -0.059 | 15 | 20 | 0.000 | 15 | 30 | 0.000 |
| 16 | 10 | -0.001 | 16 | 20 | -0.015 | 16 | 30 | -0.026 |
| 17 | 10 | 0.000 | 17 | 20 | 0.449 | 17 | 30 | -0.072 |
| 18 | 10 | 0.000 | 18 | 20 | -0.044 | 18 | 30 | 0.035 |
| 19 | 10 | -0.001 | 19 | 20 | 0.001 | 19 | 30 | -0.381 |
| 20 | 10 | 0.000 | 20 | 20 | -8.164 | 20 | 30 | -0.059 |
| 21 | 10 | 0.000 | 21 | 20 | -0.016 | 21 | 30 | 0.057 |
| 22 | 10 | 0.000 | 22 | 20 | -0.054 | 22 | 30 | 0.002 |
| 23 | 10 | -0.002 | 23 | 20 | 0.063 | 23 | 30 | 0.116 |
| 24 | 10 | 0.000 | 24 | 20 | -0.070 | 24 | 30 | 0.041 |
| 25 | 10 | -0.001 | 25 | 20 | 0.358 | 25 | 30 | 0.268 |
| 26 | 10 | 0.000 | 26 | 20 | -0.025 | 26 | 30 | -0.008 |
| 27 | 10 | 0.002 | 27 | 20 | -0.091 | 27 | 30 | -0.072 |
| 28 | 10 | 0.002 | 28 | 20 | -0.001 | 28 | 30 | -0.334 |
| 29 | 10 | -0.002 | 29 | 20 | 0.281 | 29 | 30 | -0.043 |
| 30 | 10 | 0.000 | 30 | 20 | -0.059 | 30 | 30 | -8.166 |

**Tight-binding Hamiltonian for ([$Mn_3$]$_2$-(ada)$_3$)$^{2+}$**

| | | | | | | | | |
|---|---|---|---|---|---|---|---|---|
| 1 | 1 | -4.692 | 1 | 11 | -0.121 | 1 | 21 | -0.063 |
| 2 | 1 | 0.008 | 2 | 11 | -0.036 | 2 | 21 | -0.030 |
| 3 | 1 | 0.009 | 3 | 11 | -0.051 | 3 | 21 | 0.020 |
| 4 | 1 | 0.098 | 4 | 11 | 0.362 | 4 | 21 | 0.154 |
| 5 | 1 | 0.067 | 5 | 11 | -0.039 | 5 | 21 | 0.009 |
| 6 | 1 | 0.001 | 6 | 11 | 0.001 | 6 | 21 | 0.000 |
| 7 | 1 | -0.015 | 7 | 11 | 0.001 | 7 | 21 | 0.000 |
| 8 | 1 | -0.002 | 8 | 11 | 0.000 | 8 | 21 | 0.001 |
| 9 | 1 | 0.003 | 9 | 11 | 0.001 | 9 | 21 | 0.000 |
| 10 | 1 | -0.002 | 10 | 11 | 0.000 | 10 | 21 | 0.000 |
| 11 | 1 | -0.121 | 11 | 11 | -7.860 | 11 | 21 | -0.008 |
| 12 | 1 | -0.035 | 12 | 11 | -0.521 | 12 | 21 | -0.098 |
| 13 | 1 | -0.004 | 13 | 11 | -0.082 | 13 | 21 | -0.103 |
| 14 | 1 | -0.097 | 14 | 11 | -0.259 | 14 | 21 | -0.007 |
| 15 | 1 | 0.077 | 15 | 11 | -0.074 | 15 | 21 | 0.127 |
| 16 | 1 | -0.001 | 16 | 11 | -0.001 | 16 | 21 | 0.001 |
| 17 | 1 | -0.005 | 17 | 11 | 0.001 | 17 | 21 | 0.000 |
| 18 | 1 | -0.001 | 18 | 11 | 0.000 | 18 | 21 | 0.000 |
| 19 | 1 | 0.001 | 19 | 11 | -0.001 | 19 | 21 | 0.001 |
| 20 | 1 | 0.004 | 20 | 11 | -0.001 | 20 | 21 | 0.000 |
| 21 | 1 | -0.063 | 21 | 11 | -0.008 | 21 | 21 | -8.202 |
| 22 | 1 | 0.131 | 22 | 11 | 0.065 | 22 | 21 | 0.156 |
| 23 | 1 | 0.008 | 23 | 11 | 0.186 | 23 | 21 | -0.016 |
| 24 | 1 | -0.093 | 24 | 11 | -0.005 | 24 | 21 | -0.102 |
| 25 | 1 | -0.438 | 25 | 11 | 0.257 | 25 | 21 | 0.082 |
| 26 | 1 | -0.003 | 26 | 11 | 0.000 | 26 | 21 | -0.001 |
| 27 | 1 | -0.010 | 27 | 11 | 0.000 | 27 | 21 | 0.000 |
| 28 | 1 | 0.015 | 28 | 11 | 0.000 | 28 | 21 | -0.001 |
| 29 | 1 | -0.002 | 29 | 11 | 0.000 | 29 | 21 | 0.000 |
| 30 | 1 | -0.001 | 30 | 11 | -0.002 | 30 | 21 | 0.000 |
| 1 | 2 | 0.008 | 1 | 12 | -0.035 | 1 | 22 | 0.131 |
| 2 | 2 | -2.927 | 2 | 12 | -0.001 | 2 | 22 | -0.005 |
| 3 | 2 | 0.403 | 3 | 12 | -0.022 | 3 | 22 | -0.039 |
| 4 | 2 | -0.027 | 4 | 12 | -0.015 | 4 | 22 | -0.248 |
| 5 | 2 | 0.337 | 5 | 12 | 0.021 | 5 | 22 | -0.051 |
| 6 | 2 | -0.001 | 6 | 12 | -0.001 | 6 | 22 | 0.000 |
| 7 | 2 | 0.000 | 7 | 12 | -0.001 | 7 | 22 | -0.001 |
| 8 | 2 | 0.001 | 8 | 12 | 0.000 | 8 | 22 | -0.001 |
| 9 | 2 | 0.001 | 9 | 12 | -0.001 | 9 | 22 | 0.000 |
| 10 | 2 | -0.001 | 10 | 12 | 0.000 | 10 | 22 | 0.001 |
| 11 | 2 | -0.036 | 11 | 12 | -0.521 | 11 | 22 | 0.065 |
| 12 | 2 | -0.001 | 12 | 12 | -2.930 | 12 | 22 | -0.072 |
| 13 | 2 | -0.041 | 13 | 12 | -0.026 | 13 | 22 | 0.355 |
| 14 | 2 | 0.120 | 14 | 12 | 0.037 | 14 | 22 | -0.005 |
| 15 | 2 | 0.092 | 15 | 12 | -0.004 | 15 | 22 | -0.060 |
| 16 | 2 | 0.000 | 16 | 12 | 0.000 | 16 | 22 | 0.000 |
| 17 | 2 | 0.001 | 17 | 12 | -0.001 | 17 | 22 | -0.001 |
| 18 | 2 | 0.002 | 18 | 12 | 0.000 | 18 | 22 | 0.000 |
| 19 | 2 | 0.000 | 19 | 12 | 0.000 | 19 | 22 | 0.000 |
| 20 | 2 | 0.001 | 20 | 12 | 0.000 | 20 | 22 | 0.001 |
| 21 | 2 | -0.030 | 21 | 12 | -0.098 | 21 | 22 | 0.156 |
| 22 | 2 | -0.005 | 22 | 12 | -0.072 | 22 | 22 | -7.718 |
| 23 | 2 | -0.033 | 23 | 12 | 0.097 | 23 | 22 | -0.062 |
| 24 | 2 | 0.002 | 24 | 12 | 0.002 | 24 | 22 | 0.528 |
| 25 | 2 | 0.021 | 25 | 12 | 0.018 | 25 | 22 | 0.123 |
| 26 | 2 | 0.000 | 26 | 12 | -0.001 | 26 | 22 | -0.001 |
| 27 | 2 | 0.000 | 27 | 12 | 0.000 | 27 | 22 | -0.001 |
| 28 | 2 | 0.001 | 28 | 12 | 0.001 | 28 | 22 | 0.001 |
| 29 | 2 | 0.000 | 29 | 12 | 0.001 | 29 | 22 | 0.000 |
| 30 | 2 | 0.000 | 30 | 12 | 0.000 | 30 | 22 | 0.001 |
| 1 | 3 | 0.009 | 1 | 13 | -0.004 | 1 | 23 | 0.008 |
| 2 | 3 | 0.403 | 2 | 13 | -0.041 | 2 | 23 | -0.033 |
| 3 | 3 | -8.169 | 3 | 13 | 0.274 | 3 | 23 | 0.195 |
| 4 | 3 | 0.000 | 4 | 13 | -0.249 | 4 | 23 | 0.035 |
| 5 | 3 | 0.156 | 5 | 13 | 0.120 | 5 | 23 | 0.033 |
| 6 | 3 | 0.000 | 6 | 13 | 0.003 | 6 | 23 | 0.005 |
| 7 | 3 | 0.000 | 7 | 13 | 0.001 | 7 | 23 | 0.014 |
| 8 | 3 | 0.001 | 8 | 13 | 0.001 | 8 | 23 | 0.003 |
| 9 | 3 | 0.001 | 9 | 13 | 0.003 | 9 | 23 | 0.003 |
| 10 | 3 | -0.001 | 10 | 13 | 0.007 | 10 | 23 | -0.011 |
| 11 | 3 | -0.051 | 11 | 13 | -0.082 | 11 | 23 | 0.186 |
| 12 | 3 | -0.022 | 12 | 13 | -0.026 | 12 | 23 | 0.097 |
| 13 | 3 | 0.274 | 13 | 13 | -3.274 | 13 | 23 | 0.460 |
| 14 | 3 | 0.007 | 14 | 13 | 0.122 | 14 | 23 | 0.026 |
| 15 | 3 | -0.105 | 15 | 13 | 0.197 | 15 | 23 | 0.109 |
| 16 | 3 | 0.001 | 16 | 13 | 0.000 | 16 | 23 | -0.001 |
| 17 | 3 | 0.001 | 17 | 13 | 0.000 | 17 | 23 | 0.013 |
| 18 | 3 | 0.000 | 18 | 13 | 0.003 | 18 | 23 | 0.000 |
| 19 | 3 | 0.000 | 19 | 13 | -0.003 | 19 | 23 | 0.005 |
| 20 | 3 | 0.000 | 20 | 13 | 0.007 | 20 | 23 | -0.005 |
| 21 | 3 | 0.020 | 21 | 13 | -0.103 | 21 | 23 | -0.016 |
| 22 | 3 | -0.039 | 22 | 13 | 0.355 | 22 | 23 | -0.062 |
| 23 | 3 | 0.195 | 23 | 13 | 0.460 | 23 | 23 | -4.680 |
| 24 | 3 | 0.042 | 24 | 13 | 0.012 | 24 | 23 | -0.006 |
| 25 | 3 | 0.249 | 25 | 13 | 0.151 | 25 | 23 | 0.126 |
| 26 | 3 | -0.001 | 26 | 13 | 0.003 | 26 | 23 | 0.001 |
| 27 | 3 | 0.000 | 27 | 13 | 0.005 | 27 | 23 | -0.002 |
| 28 | 3 | 0.001 | 28 | 13 | -0.002 | 28 | 23 | 0.004 |
| 29 | 3 | 0.000 | 29 | 13 | -0.003 | 29 | 23 | 0.001 |
| 30 | 3 | 0.000 | 30 | 13 | 0.003 | 30 | 23 | -0.005 |
| 1 | 4 | 0.098 | 1 | 14 | -0.097 | 1 | 24 | -0.093 |
| 2 | 4 | -0.027 | 2 | 14 | 0.120 | 2 | 24 | 0.002 |
| 3 | 4 | 0.000 | 3 | 14 | 0.007 | 3 | 24 | 0.042 |
| 4 | 4 | -3.255 | 4 | 14 | -0.028 | 4 | 24 | 0.035 |
| 5 | 4 | 0.149 | 5 | 14 | 0.014 | 5 | 24 | -0.119 |
| 6 | 4 | 0.001 | 6 | 14 | 0.000 | 6 | 24 | 0.000 |
| 7 | 4 | -0.003 | 7 | 14 | 0.001 | 7 | 24 | -0.001 |
| 8 | 4 | -0.004 | 8 | 14 | -0.001 | 8 | 24 | 0.000 |
| 9 | 4 | 0.002 | 9 | 14 | 0.000 | 9 | 24 | 0.000 |
| 10 | 4 | 0.005 | 10 | 14 | 0.001 | 10 | 24 | 0.000 |
| 11 | 4 | 0.362 | 11 | 14 | -0.259 | 11 | 24 | -0.005 |
| 12 | 4 | -0.015 | 12 | 14 | 0.037 | 12 | 24 | 0.002 |
| 13 | 4 | -0.249 | 13 | 14 | 0.122 | 13 | 24 | 0.012 |
| 14 | 4 | -0.028 | 14 | 14 | -8.032 | 14 | 24 | -0.030 |
| 15 | 4 | 0.420 | 15 | 14 | 0.054 | 15 | 24 | 0.034 |
| 16 | 4 | 0.001 | 16 | 14 | -0.001 | 16 | 24 | -0.001 |
| 17 | 4 | 0.000 | 17 | 14 | 0.000 | 17 | 24 | 0.001 |
| 18 | 4 | -0.003 | 18 | 14 | 0.000 | 18 | 24 | -0.001 |
| 19 | 4 | -0.008 | 19 | 14 | 0.000 | 19 | 24 | -0.001 |
| 20 | 4 | -0.001 | 20 | 14 | 0.000 | 20 | 24 | 0.001 |
| 21 | 4 | 0.154 | 21 | 14 | -0.007 | 21 | 24 | -0.102 |
| 22 | 4 | -0.248 | 22 | 14 | -0.005 | 22 | 24 | 0.528 |
| 23 | 4 | 0.035 | 23 | 14 | 0.026 | 23 | 24 | -0.006 |
| 24 | 4 | 0.035 | 24 | 14 | -0.030 | 24 | 24 | -2.961 |
| 25 | 4 | 0.217 | 25 | 14 | 0.061 | 25 | 24 | -0.027 |
| 26 | 4 | 0.000 | 26 | 14 | 0.002 | 26 | 24 | 0.000 |
| 27 | 4 | 0.004 | 27 | 14 | 0.001 | 27 | 24 | 0.000 |
| 28 | 4 | -0.001 | 28 | 14 | 0.000 | 28 | 24 | 0.001 |
| 29 | 4 | 0.001 | 29 | 14 | 0.000 | 29 | 24 | -0.002 |
| 30 | 4 | -0.007 | 30 | 14 | 0.000 | 30 | 24 | 0.000 |
| 1 | 5 | 0.067 | 1 | 15 | 0.077 | 1 | 25 | -0.438 |
| 2 | 5 | 0.337 | 2 | 15 | 0.092 | 2 | 25 | 0.021 |
| 3 | 5 | 0.156 | 3 | 15 | -0.105 | 3 | 25 | 0.249 |
| 4 | 5 | 0.149 | 4 | 15 | 0.420 | 4 | 25 | 0.217 |
| 5 | 5 | -7.729 | 5 | 15 | -0.064 | 5 | 25 | 0.235 |
| 6 | 5 | 0.000 | 6 | 15 | 0.000 | 6 | 25 | 0.003 |
| 7 | 5 | -0.001 | 7 | 15 | 0.005 | 7 | 25 | -0.004 |
| 8 | 5 | -0.001 | 8 | 15 | -0.001 | 8 | 25 | -0.001 |
| 9 | 5 | 0.001 | 9 | 15 | -0.003 | 9 | 25 | 0.005 |
| 10 | 5 | 0.000 | 10 | 15 | -0.001 | 10 | 25 | -0.002 |
| 11 | 5 | -0.039 | 11 | 15 | -0.074 | 11 | 25 | 0.257 |
| 12 | 5 | 0.021 | 12 | 15 | -0.004 | 12 | 25 | 0.018 |
| 13 | 5 | 0.120 | 13 | 15 | 0.197 | 13 | 25 | 0.151 |
| 14 | 5 | 0.014 | 14 | 15 | 0.054 | 14 | 25 | 0.061 |
| 15 | 5 | -0.064 | 15 | 15 | -4.659 | 15 | 25 | -0.004 |
| 16 | 5 | -0.001 | 16 | 15 | 0.003 | 16 | 25 | 0.004 |
| 17 | 5 | 0.000 | 17 | 15 | -0.016 | 17 | 25 | -0.006 |
| 18 | 5 | 0.000 | 18 | 15 | -0.005 | 18 | 25 | 0.001 |
| 19 | 5 | -0.001 | 19 | 15 | 0.007 | 19 | 25 | 0.004 |
| 20 | 5 | 0.001 | 20 | 15 | -0.001 | 20 | 25 | 0.003 |
| 21 | 5 | 0.009 | 21 | 15 | 0.127 | 21 | 25 | 0.082 |
| 22 | 5 | -0.051 | 22 | 15 | -0.060 | 22 | 25 | 0.123 |
| 23 | 5 | 0.033 | 23 | 15 | 0.109 | 23 | 25 | 0.126 |
| 24 | 5 | -0.119 | 24 | 15 | 0.034 | 24 | 25 | -0.027 |
| 25 | 5 | 0.235 | 25 | 15 | -0.004 | 25 | 25 | -3.263 |
| 26 | 5 | 0.000 | 26 | 15 | 0.002 | 26 | 25 | 0.001 |
| 27 | 5 | 0.000 | 27 | 15 | -0.003 | 27 | 25 | -0.006 |
| 28 | 5 | 0.001 | 28 | 15 | -0.013 | 28 | 25 | -0.002 |
| 29 | 5 | 0.000 | 29 | 15 | 0.001 | 29 | 25 | 0.006 |
| 30 | 5 | -0.001 | 30 | 15 | 0.008 | 30 | 25 | 0.000 |
| 1 | 6 | 0.001 | 1 | 16 | -0.001 | 1 | 26 | -0.003 |
| 2 | 6 | -0.001 | 2 | 16 | 0.000 | 2 | 26 | 0.000 |
| 3 | 6 | 0.000 | 3 | 16 | 0.001 | 3 | 26 | -0.001 |
| 4 | 6 | 0.001 | 4 | 16 | 0.001 | 4 | 26 | 0.000 |
| 5 | 6 | 0.000 | 5 | 16 | -0.001 | 5 | 26 | 0.000 |
| 6 | 6 | -3.035 | 6 | 16 | -0.007 | 6 | 26 | 0.008 |
| 7 | 6 | 0.250 | 7 | 16 | 0.019 | 7 | 26 | 0.076 |
| 8 | 6 | 0.052 | 8 | 16 | 0.012 | 8 | 26 | -0.012 |
| 9 | 6 | -0.266 | 9 | 16 | 0.001 | 9 | 26 | 0.007 |
| 10 | 6 | 0.235 | 10 | 16 | 0.025 | 10 | 26 | 0.041 |
| 11 | 6 | 0.001 | 11 | 16 | -0.001 | 11 | 26 | 0.000 |
| 12 | 6 | -0.001 | 12 | 16 | 0.000 | 12 | 26 | -0.001 |
| 13 | 6 | 0.003 | 13 | 16 | 0.000 | 13 | 26 | 0.003 |
| 14 | 6 | 0.000 | 14 | 16 | -0.001 | 14 | 26 | 0.002 |
| 15 | 6 | 0.000 | 15 | 16 | 0.003 | 15 | 26 | 0.002 |
| 16 | 6 | -0.007 | 16 | 16 | -3.184 | 16 | 26 | -0.010 |
| 17 | 6 | 0.049 | 17 | 16 | -0.070 | 17 | 26 | 0.014 |
| 18 | 6 | -0.094 | 18 | 16 | -0.060 | 18 | 26 | -0.009 |
| 19 | 6 | 0.018 | 19 | 16 | -0.158 | 19 | 26 | 0.026 |
| 20 | 6 | -0.024 | 20 | 16 | -0.017 | 20 | 26 | -0.005 |
| 21 | 6 | 0.000 | 21 | 16 | 0.001 | 21 | 26 | -0.001 |
| 22 | 6 | 0.000 | 22 | 16 | 0.000 | 22 | 26 | -0.001 |
| 23 | 6 | 0.005 | 23 | 16 | -0.001 | 23 | 26 | 0.001 |
| 24 | 6 | 0.000 | 24 | 16 | -0.001 | 24 | 26 | 0.000 |
| 25 | 6 | 0.003 | 25 | 16 | 0.004 | 25 | 26 | 0.001 |
| 26 | 6 | 0.008 | 26 | 16 | -0.010 | 26 | 26 | -3.196 |
| 27 | 6 | -0.040 | 27 | 16 | -0.043 | 27 | 26 | -0.134 |
| 28 | 6 | 0.051 | 28 | 16 | -0.066 | 28 | 26 | -0.090 |
| 29 | 6 | -0.088 | 29 | 16 | 0.028 | 29 | 26 | 0.115 |
| 30 | 6 | 0.035 | 30 | 16 | 0.023 | 30 | 26 | -0.009 |
| 1 | 7 | -0.015 | 1 | 17 | -0.005 | 1 | 27 | -0.010 |
| 2 | 7 | 0.000 | 2 | 17 | 0.001 | 2 | 27 | 0.000 |
| 3 | 7 | 0.000 | 3 | 17 | 0.001 | 3 | 27 | 0.000 |
| 4 | 7 | -0.003 | 4 | 17 | 0.000 | 4 | 27 | 0.004 |
| 5 | 7 | -0.001 | 5 | 17 | 0.000 | 5 | 27 | 0.000 |
| 6 | 7 | 0.250 | 6 | 17 | 0.049 | 6 | 27 | -0.040 |
| 7 | 7 | -2.267 | 7 | 17 | -0.012 | 7 | 27 | -0.004 |
| 8 | 7 | 0.093 | 8 | 17 | 0.068 | 8 | 27 | 0.028 |
| 9 | 7 | 0.187 | 9 | 17 | -0.062 | 9 | 27 | -0.061 |
| 10 | 7 | -0.208 | 10 | 17 | 0.038 | 10 | 27 | -0.028 |
| 11 | 7 | 0.001 | 11 | 17 | 0.001 | 11 | 27 | 0.000 |
| 12 | 7 | -0.001 | 12 | 17 | -0.001 | 12 | 27 | 0.000 |
| 13 | 7 | 0.001 | 13 | 17 | 0.000 | 13 | 27 | 0.005 |
| 14 | 7 | 0.001 | 14 | 17 | 0.000 | 14 | 27 | 0.001 |
| 15 | 7 | 0.005 | 15 | 17 | -0.016 | 15 | 27 | -0.003 |
| 16 | 7 | 0.019 | 16 | 17 | -0.070 | 16 | 27 | -0.043 |
| 17 | 7 | -0.012 | 17 | 17 | -2.191 | 17 | 27 | 0.077 |
| 18 | 7 | -0.040 | 18 | 17 | 0.203 | 18 | 27 | 0.055 |
| 19 | 7 | 0.067 | 19 | 17 | -0.161 | 19 | 27 | 0.045 |
| 20 | 7 | 0.013 | 20 | 17 | 0.097 | 20 | 27 | 0.051 |
| 21 | 7 | 0.000 | 21 | 17 | 0.000 | 21 | 27 | 0.000 |
| 22 | 7 | -0.001 | 22 | 17 | -0.001 | 22 | 27 | -0.001 |
| 23 | 7 | 0.014 | 23 | 17 | 0.013 | 23 | 27 | -0.002 |
| 24 | 7 | -0.001 | 24 | 17 | 0.001 | 24 | 27 | 0.000 |
| 25 | 7 | -0.004 | 25 | 17 | -0.006 | 25 | 27 | -0.006 |
| 26 | 7 | 0.076 | 26 | 17 | 0.014 | 26 | 27 | -0.134 |
| 27 | 7 | -0.004 | 27 | 17 | 0.077 | 27 | 27 | -2.599 |
| 28 | 7 | 0.025 | 28 | 17 | -0.024 | 28 | 27 | -0.189 |
| 29 | 7 | 0.055 | 29 | 17 | 0.019 | 29 | 27 | -0.349 |
| 30 | 7 | 0.038 | 30 | 17 | -0.006 | 30 | 27 | 0.239 |
| 1 | 8 | -0.002 | 1 | 18 | -0.001 | 1 | 28 | 0.015 |
| 2 | 8 | 0.001 | 2 | 18 | 0.002 | 2 | 28 | 0.001 |
| 3 | 8 | 0.001 | 3 | 18 | 0.000 | 3 | 28 | 0.001 |
| 4 | 8 | -0.004 | 4 | 18 | -0.003 | 4 | 28 | -0.001 |
| 5 | 8 | -0.001 | 5 | 18 | 0.000 | 5 | 28 | 0.001 |
| 6 | 8 | 0.052 | 6 | 18 | -0.094 | 6 | 28 | 0.051 |
| 7 | 8 | 0.093 | 7 | 18 | -0.040 | 7 | 28 | 0.025 |
| 8 | 8 | -3.185 | 8 | 18 | 0.012 | 8 | 28 | 0.015 |
| 9 | 8 | 0.047 | 9 | 18 | -0.037 | 9 | 28 | 0.017 |
| 10 | 8 | 0.127 | 10 | 18 | 0.018 | 10 | 28 | -0.079 |
| 11 | 8 | 0.000 | 11 | 18 | 0.000 | 11 | 28 | 0.000 |
| 12 | 8 | 0.000 | 12 | 18 | 0.000 | 12 | 28 | 0.001 |
| 13 | 8 | 0.001 | 13 | 18 | 0.003 | 13 | 28 | -0.002 |
| 14 | 8 | -0.001 | 14 | 18 | 0.000 | 14 | 28 | 0.000 |
| 15 | 8 | -0.001 | 15 | 18 | -0.005 | 15 | 28 | -0.013 |
| 16 | 8 | 0.012 | 16 | 18 | -0.060 | 16 | 28 | -0.066 |
| 17 | 8 | 0.068 | 17 | 18 | 0.203 | 17 | 28 | -0.024 |
| 18 | 8 | 0.012 | 18 | 18 | -2.980 | 18 | 28 | 0.052 |
| 19 | 8 | 0.051 | 19 | 18 | 0.224 | 19 | 28 | 0.036 |
| 20 | 8 | -0.011 | 20 | 18 | -0.319 | 20 | 28 | -0.081 |
| 21 | 8 | 0.001 | 21 | 18 | 0.000 | 21 | 28 | -0.001 |
| 22 | 8 | -0.001 | 22 | 18 | 0.000 | 22 | 28 | 0.001 |
| 23 | 8 | 0.003 | 23 | 18 | 0.000 | 23 | 28 | 0.004 |
| 24 | 8 | 0.000 | 24 | 18 | -0.001 | 24 | 28 | 0.001 |
| 25 | 8 | -0.001 | 25 | 18 | 0.001 | 25 | 28 | -0.002 |
| 26 | 8 | -0.012 | 26 | 18 | -0.009 | 26 | 28 | -0.090 |
| 27 | 8 | 0.028 | 27 | 18 | 0.055 | 27 | 28 | -0.189 |
| 28 | 8 | 0.015 | 28 | 18 | 0.052 | 28 | 28 | -2.210 |
| 29 | 8 | -0.005 | 29 | 18 | 0.082 | 29 | 28 | -0.091 |
| 30 | 8 | -0.009 | 30 | 18 | -0.029 | 30 | 28 | -0.095 |
| 1 | 9 | 0.003 | 1 | 19 | 0.001 | 1 | 29 | -0.002 |
| 2 | 9 | 0.001 | 2 | 19 | 0.000 | 2 | 29 | 0.000 |
| 3 | 9 | 0.001 | 3 | 19 | 0.000 | 3 | 29 | 0.000 |
| 4 | 9 | 0.002 | 4 | 19 | -0.008 | 4 | 29 | 0.001 |
| 5 | 9 | 0.001 | 5 | 19 | -0.001 | 5 | 29 | 0.000 |
| 6 | 9 | -0.266 | 6 | 19 | 0.018 | 6 | 29 | -0.088 |
| 7 | 9 | 0.187 | 7 | 19 | 0.067 | 7 | 29 | 0.055 |
| 8 | 9 | 0.047 | 8 | 19 | 0.051 | 8 | 29 | -0.005 |
| 9 | 9 | -3.124 | 9 | 19 | 0.052 | 9 | 29 | 0.021 |
| 10 | 9 | 0.157 | 10 | 19 | 0.033 | 10 | 29 | 0.016 |
| 11 | 9 | 0.001 | 11 | 19 | -0.001 | 11 | 29 | 0.000 |
| 12 | 9 | -0.001 | 12 | 19 | 0.000 | 12 | 29 | 0.001 |
| 13 | 9 | 0.003 | 13 | 19 | -0.003 | 13 | 29 | -0.003 |
| 14 | 9 | 0.000 | 14 | 19 | 0.000 | 14 | 29 | 0.000 |
| 15 | 9 | -0.003 | 15 | 19 | 0.007 | 15 | 29 | 0.001 |
| 16 | 9 | 0.001 | 16 | 19 | -0.158 | 16 | 29 | 0.028 |
| 17 | 9 | -0.062 | 17 | 19 | -0.161 | 17 | 29 | 0.019 |
| 18 | 9 | -0.037 | 18 | 19 | 0.224 | 18 | 29 | 0.082 |
| 19 | 9 | 0.052 | 19 | 19 | -2.391 | 19 | 29 | -0.018 |
| 20 | 9 | -0.089 | 20 | 19 | 0.081 | 20 | 29 | 0.015 |
| 21 | 9 | 0.000 | 21 | 19 | 0.001 | 21 | 29 | 0.000 |
| 22 | 9 | 0.000 | 22 | 19 | 0.000 | 22 | 29 | 0.000 |
| 23 | 9 | 0.003 | 23 | 19 | 0.005 | 23 | 29 | 0.001 |
| 24 | 9 | 0.000 | 24 | 19 | -0.001 | 24 | 29 | -0.002 |
| 25 | 9 | 0.005 | 25 | 19 | 0.004 | 25 | 29 | 0.006 |
| 26 | 9 | 0.007 | 26 | 19 | 0.026 | 26 | 29 | 0.115 |
| 27 | 9 | -0.061 | 27 | 19 | 0.045 | 27 | 29 | -0.349 |
| 28 | 9 | 0.017 | 28 | 19 | 0.036 | 28 | 29 | -0.091 |
| 29 | 9 | 0.021 | 29 | 19 | -0.018 | 29 | 29 | -2.756 |
| 30 | 9 | 0.095 | 30 | 19 | 0.053 | 30 | 29 | 0.271 |
| 1 | 10 | -0.002 | 1 | 20 | 0.004 | 1 | 30 | -0.001 |
| 2 | 10 | -0.001 | 2 | 20 | 0.001 | 2 | 30 | 0.000 |
| 3 | 10 | -0.001 | 3 | 20 | 0.000 | 3 | 30 | 0.000 |
| 4 | 10 | 0.005 | 4 | 20 | -0.001 | 4 | 30 | -0.007 |
| 5 | 10 | 0.000 | 5 | 20 | 0.001 | 5 | 30 | -0.001 |
| 6 | 10 | 0.235 | 6 | 20 | -0.024 | 6 | 30 | 0.035 |
| 7 | 10 | -0.208 | 7 | 20 | 0.013 | 7 | 30 | 0.038 |
| 8 | 10 | 0.127 | 8 | 20 | -0.011 | 8 | 30 | -0.009 |
| 9 | 10 | 0.157 | 9 | 20 | -0.089 | 9 | 30 | 0.095 |
| 10 | 10 | -2.373 | 10 | 20 | 0.074 | 10 | 30 | -0.050 |
| 11 | 10 | 0.000 | 11 | 20 | -0.001 | 11 | 30 | -0.002 |
| 12 | 10 | 0.000 | 12 | 20 | 0.000 | 12 | 30 | 0.000 |
| 13 | 10 | 0.007 | 13 | 20 | 0.007 | 13 | 30 | 0.003 |
| 14 | 10 | 0.001 | 14 | 20 | 0.000 | 14 | 30 | 0.000 |
| 15 | 10 | -0.001 | 15 | 20 | -0.001 | 15 | 30 | 0.008 |
| 16 | 10 | 0.025 | 16 | 20 | -0.017 | 16 | 30 | 0.023 |
| 17 | 10 | 0.038 | 17 | 20 | 0.097 | 17 | 30 | -0.006 |
| 18 | 10 | 0.018 | 18 | 20 | -0.319 | 18 | 30 | -0.029 |
| 19 | 10 | 0.033 | 19 | 20 | 0.081 | 19 | 30 | 0.053 |
| 20 | 10 | 0.074 | 20 | 20 | -3.228 | 20 | 30 | -0.089 |
| 21 | 10 | 0.000 | 21 | 20 | 0.000 | 21 | 30 | 0.000 |
| 22 | 10 | 0.001 | 22 | 20 | 0.001 | 22 | 30 | 0.001 |
| 23 | 10 | -0.011 | 23 | 20 | -0.005 | 23 | 30 | -0.005 |
| 24 | 10 | 0.000 | 24 | 20 | 0.001 | 24 | 30 | 0.000 |
| 25 | 10 | -0.002 | 25 | 20 | 0.003 | 25 | 30 | 0.000 |
| 26 | 10 | 0.041 | 26 | 20 | -0.005 | 26 | 30 | -0.009 |
| 27 | 10 | -0.028 | 27 | 20 | 0.051 | 27 | 30 | 0.239 |
| 28 | 10 | -0.079 | 28 | 20 | -0.081 | 28 | 30 | -0.095 |
| 29 | 10 | 0.016 | 29 | 20 | 0.015 | 29 | 30 | 0.271 |
| 30 | 10 | -0.050 | 30 | 20 | -0.089 | 30 | 30 | -3.261 |